 \documentclass{aa}
\usepackage{graphicx}
\usepackage{url}
\usepackage{natbib}
\begin{document}
   \title{The XMM-LSS survey:}
   \subtitle{The XMDS/VVDS $4 \sigma$ catalogue
    \thanks{The full catalogue illustrated in Table~\ref{TabCat} is only available in electronic form
            at the CDS via anonymous ftp to {\tt cdsarc.u-strasbg.fr (130.79.128.5)}
            or via {\tt http://cdsweb.u-strasbg.fr/cgi-bin/qcat?J/A+A/}} }
   \titlerunning{The XMDS/VVDS $4 \sigma$ catalogue}

   \author{L.Chiappetti   \inst{1} \and
           M.Tajer        \inst{2,5} \and
           G.Trinchieri   \inst{2} \and
           D.Maccagni     \inst{1} \and
           L.Maraschi     \inst{2} \and
           L.Paioro       \inst{1} \and
           M.Pierre       \inst{3} \and
           J.Surdej       \inst{4} \and
           O.Garcet       \inst{4} \and
           E.Gosset       \inst{4} \and
           O.Le F{\` e}vre \inst{6} \and
           E.Bertin       \inst{7} \and
           H.J.McCracken  \inst{7} \and
           Y.Mellier      \inst{7} \and
           S.Foucaud      \inst{1} \and
           M.Radovich     \inst{8} \and
           V.Ripepi       \inst{8} \and
           M.Arnaboldi    \inst{9}
          }

%  \offprints{L.Chiappetti {\email {lucio@mi.iasf.cnr.it}}}

   \institute{ INAF IASF,
               Sezione di Milano ``G.Occhialini'',
               via Bassini 15, I-20133 Milano, Italy \\
               {\email {lucio@mi.iasf.cnr.it}}
         \and
              INAF Osservatorio di Brera,
              via Brera 28, I-20121 Milano, Italy
         \and
              CEA/DSM/DAPNIA
              Service d'Astrophysique, Saclay,
              F-91191 Gif sur Yvette, France
         \and
              Institut d'Astrophysique et de G\'eophysique,
              Universit\'e de Li\`ege,
              All\'ee du 6 Ao\^ut 17, B-4000 Li\`ege 1, Belgium
         \and
              Universit\`a degli Studi di Milano - Bicocca, 
              Dipartimento di Fisica, 
              Piazza della Scienza 3, I-20126 Milano, Italy
         \and 
              Laboratoire d'Astrophysique de Marseille,
              Traverse du Siphon, F-13376 Marseille, France
         \and
              Institut d'Astrophysique de Paris,
              89bis bvd Arago, F-75014 Paris, France
         \and
              INAF Osservatorio di Capodimonte,
              via Moiariello 16, I-80131 Napoli, Italy
         \and
              INAF Osservatorio Astronomico di Torino,
              via Osservatorio 20, I-10025 Pino Torinese, Italy
         }

   \date{Received <date> / accepted <date>}

   \abstract{ We present a first catalogue of X-ray sources resulting from the central
              area of the XMM-LSS (Large Scale Structure survey). We describe the
              reduction procedures and the database tools we developed and used to derive
              a well defined catalogue of X-ray sources. The present catalogue is 
              limited to a sub-sample of 286 sources detected at $4 \sigma$ in the 
              1 deg$^2$ area covered by the photometric
              VVDS (VIRMOS VLT Deep Survey), which allows us to provide optical and
              radio identifications. We also discuss
              the X-ray properties of a larger X-ray sample of 536 sources detected at $> 4
              \sigma$ in the full 3 deg$^2$ area  of the \textit{XMM} Medium Deep Survey (XMDS)
              independently of the optical identification. We also derive
              the logN--logS relationship for a sample of more than one thousand sources that we
              discuss in the context of other surveys at similar fluxes.
    }

   \maketitle

\section{Introduction \label{SecIntro}}

The XMDS (\textit{XMM} Medium Deep Survey) is an X-ray survey 
based on the pooling of a significant fraction
of the \textit{XMM-Newton} guaranteed time of three ``hardware institutes"
(IASF Milano for XMM-EPIC, Li\`ege for XMM-OM and CEA Saclay for XMM-SSC).

The XMDS pointings lie at the heart of the full, larger \textit{XMM} Large Scale Structure
(XMM-LSS) Survey \citep[see][]{pierre04}, to which we refer for a discussion
of the motivations and the choice of the sky field.

About two thirds of the XMDS area are covered in the optical band by the
VVDS (VIRMOS VLT Deep Survey) both by UBVRI photometry 
\citep{2004A&A...417..839L} and by multi-object spectroscopy with VIMOS
\citep{2004astro.ph..9133F}, by an associated radio survey at 1.4 GHz \citep{2003A&A...403..857B}
and by a GALEX Early Release Observation \citep{2005ApJ...619L..43A}.

The XMM-LSS area has been covered by an associated radio survey at 74 and 325 MHz
\citep{2003ApJ...591..640C} and will be covered by surveys in other bands like
SWIRE \citep{2003PASP..115..897L}, the CFHTLS (see web site\footnote{\url{
http://www.cfht.hawaii.edu/Science/CFHLS/}}) and by our own follow up observations
\citep{pierre04}.

Thus, for these areas, multiwavelength information exists or will
be gathered in the future, allowing investigations of large scale
structures and source populations (clusters, galaxies and AGN) at
medium redshifts. 

For internal usage within the XMDS and XMM-LSS consortia, and in the future also
for public access, a database has been designed, which
is presented in this paper together with the first 
results obtained from the analysis of the XMDS data.

The layout of the paper is as follows. The database tool
devised to store the results of the XMM-LSS and XMDS and associated
surveys is described in Section \ref{SecDBCat}. Section \ref{SecDR} contains the analysis 
procedure \citep[closely following that of][]{2002ApJ...564..190B} used
to generate source lists from the 19 \textit{XMM} pointings of the
XMDS using a threshold that ensures a small number of spurious sources. Different selection
criteria are then applied to obtain samples used for different
purposes. A relatively low probability threshold is used to compute the logN--logS 
relation presented and discussed in Section \ref{SecLNLS}.
The analysis of the X-ray characteristics is restricted to sources with a signal
to noise ratio larger than 4, and the results are presented
in Section \ref{SecTot}. Only a fraction of the XMDS area is covered by the
VVDS survey. For the latter area the identification procedure
based on the photometric VVDS data and the derived optical
vs X-ray properties are described in Section \ref{Sec4S}. 

Preliminary accounts of this work have been presented at recent conferences
\citep{cozumel03,volterra04}.

\section{The database and catalogue \label{SecDBCat}}
\subsection{The LSS database \label{SecDB}}

The web site of the XMM-LSS Master Catalogue\footnote{\url{
http://cosmos.mi.iasf.cnr.it/~lssadmin/Website/LSS}} has been designed 
as a front end to access the database containing the catalogues, both
as a working tool for the XMDS and XMM-LSS consortia and, in the near future,
for public access.

The query interface uses a Java servlet 
(communicating via JDBC protocol to the underlying
MySQL\footnote{MySQL is an open source database server developed by
MySQL AB (see \url{http://www.mysql.com})} database)
to manage the permissions of 
different groups of people to access different (parts of) database tables,
to perform selections into multiple tables in an user-friendly way, to have
a quick look at the data and to export results as ASCII or FITS files.

The database presently includes {\em database tables} for the results of
the X-ray pipelines,
VVDS 
photometry for objects within a 40\arcsec ~box around an X-ray source,
radio and other catalogues
as reported in Section \ref{SecId}; it also provides links to
{\em data products} like
X-ray images and exposure maps, 
finding charts in the I band, radio maps, etc. 

The database includes also pre-calculated correlation tables, which link
the sequence identifiers of objects in
an X-ray table with their neighbours in other tables according to predefined
proximity criteria (e.g. a 40\arcsec ~box or a 6\arcsec ~radius), and greatly speed up
queries which involve a couple of tables.

Correlations between more than two tables are managed similarly by multi-column
correlation tables generated by our identification procedure
described below in section \ref{SecId}. They are also used to build {\em virtual tables}, i.e. simultaneous
views of columns taken from different individual tables, or even of results of
algebraic operations between different columns.

Usage of appropriate MySQL syntax during the composition of a query
allows the production  of results of arbitrary complexity
as output from our database, like e.g. {\tt ds9}
region files, or the \LaTeX ~code for the catalogue table reported here (Table~\ref{TabCat}).

\subsection{The catalogue \label{SecCat}}

In this paper we publish, as a first part of the XMDS catalogue, the results of
the identification procedure described in section \ref{SecId}, i.e. a catalogue
of 286 X-ray sources detected at more than $4 \sigma$ and located in the VVDS area.
Further releases will also include sources already present in our
database and detected at a lower S/N and/or located outside the VVDS area.

Our current choice allows us to present a catalogue inclusive of reliable
(though not spectroscopically confirmed) identifications and of the basic optical
and radio characteristics of the X-ray sources.

The catalogue published electronically at CDS along with the present paper
(and available as Online material)
includes a rather reduced, manageable number of columns. A printout sample of the catalogue
in tabular form is presented in Table~\ref{TabCat}.
Comments on individual sources are presented in Appendix \ref{SecCom}.

Since our database contains much more information than what is included in the
published catalogue, we plan
to open access to our database query engine,
in accordance with the "legacy policy" of the XMM-LSS survey.

Usage of our database interface will allow the user to make selections on the
datasets, to access additional columns (including e.g. X-ray fluxes in all
bands, photometry in additional bands, radio fluxes, identification ranks and flags,
technical information, etc.)
which is not practical to include here, and to generate (and operate upon)
expressions involving columns.

The query interface will be accessible from the time the paper is accepted at
\url{http://cosmos.mi.iasf.cnr.it/~lssadmin/Website/LSS/Query}
logging in as user {\tt xmds}, password {\tt guest} in workspace {\tt public}.

\section{Observations and data reduction \label{SecDR}}
\subsection{Observations \label{SecObs}}

\begin{figure}
   \centering
   \includegraphics[width=8.5cm]{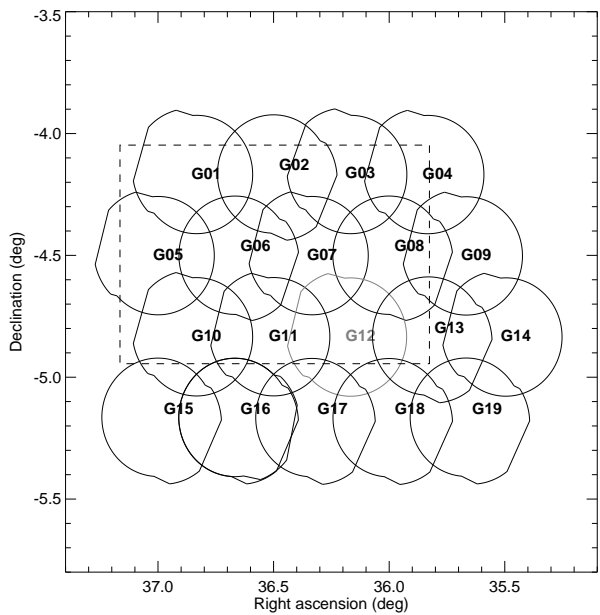}
   \caption{Location of the XMDS pointings on the sky : the approximate FOV footprint
            (with the pn protruding at the actual roll angle) is shown (solid lines);
             the gray colour for field G12 indicates that it has not
             been used (see text).
             The dashed rectangle indicates the area of VVDS photometry.
             }
    \label{FigFOV}
    \end{figure}

The XMDS observations consist of 19 overlapping pointings (typical duration in the range 20--25 ks),
covering a contiguous area of about 3 deg$^2$. 
The sky location of the combined field of view (FOV) of the pointings
is shown in Fig.~\ref{FigFOV}.

The observations were
performed by \textit{XMM-Newton} \citep{2001A&A...365L...1J},
with the EPIC MOS \citep{2001A&A...365L..27T} and pn \citep{2001A&A...365L..18S} cameras,
between July 2001 and January 2003.
A basic journal of observations is reported in Table~\ref{TabJou}, containing
the ESA dataset identifier that can be used
to inquire for further details at the ESA \textit{XMM} archive.

The XMDS fields are surrounded at present by 32 additional shorter
(typically 10 ks) pointings (B01 to B32)
performed in AO1 and AO2 as part of the XMM-LSS program. More detailed information is
available on an ancillary page of the XMM-LSS Master Catalogue web site\footnote{\url{
http://cosmos.mi.iasf.cnr.it/~lssadmin/Website/LSS/Anc/goepic.html}}.

%__________________________________________________ Two column table
   \begin{table*}
   \begin{center}
      \caption[]{Journal of XMDS observations$^e$}
         \label{TabJou}
         \begin{tabular}{lllllrrr}
            \hline
            \noalign{\smallskip}
            Field  & ESA obs id           & date & pointing RA & DEC & max exposure$^c$ & shift$^d$ $\Delta$RA & $\Delta$DEC \\
                   &                      &      &             &     & (ks)         & (\arcsec)        & (\arcsec)   \\
            \noalign{\smallskip}
            \hline
            \noalign{\smallskip}
            G01 & 011268~0101 & 28 Jan 2002 & 02:27:25.4 & --04:11:06.4 & 24.9 & --1.60 &  0.00 \\
            G02 & 011268~0201 & 14 Jul 2002 & 02:25:54.2 & --04:09:05.6 &  9.8 &  0.00 & --0.53 \\
            G03 & 011268~0301 & 19 Jan 2003 & 02:24:45.6 & --04:11:00.8 & 21.8 & --0.53 &  1.07 \\
            G04 & 010952~0101 & 29 Jan 2002 & 02:23:25.3 & --04:11:07.6 & 25.8 & --1.60 &  0.00 \\
            G05 & 011268~0401 & 02 Feb 2002 & 02:28:05.1 & --04:31:08.1 & 23.7  &  0.00 &  0.00 \\
            G06 & 011268~1301 & 26 Jul 2002 & 02:26:34.4 & --04:29:00.8 & 13.0 &  0.00 &  0.53 \\
            G07 & 011268~1001 & 30 Jan 2002 & 02:25:25.3 & --04:31:07.1 & 23.5 & --0.53 & --0.53 \\
            G08 & 011268~0501 & 25 Jul 2002 & 02:23:54.6 & --04:29:00.1 & 18.3 &  0.00 &  0.00 \\
            G09 & 010952~0601 & 31 Jan 2002 & 02:22:45.2 & --04:31:11.1 & 22.4 &  n/a  &  n/a  \\
            G10 & 010952~0201 & 29 Jan 2002 & 02:27:25.4 & --04:51:04.4 & 24.8 & --0.54 &  0.53 \\
            G11 & 010952~0301 & 02 Feb 2002 & 02:26:05.1 & --04:51:06.1 & 21.9 &  0.00 &  0.53 \\
            G12$^a$ & 010952~0401 & 01 Feb 2002 & 02:24:45.4 & --04:51:11.2 &  n/a &  n/a  &  n/a  \\
            G13 & 010952~0501 & 03 Jul 2001 & 02:23:13.1 & --04:49:03.1 & 23.9 & --2.67 & --2.13 \\
            G14 & 011268~0801 & 31 Jan 2002 & 02:22:04.1 & --04:51:09.7 & 13.6 &  n/a  &  n/a  \\
            G15 & 011111~0101 & 06 Jul 2001 & 02:27:54.1 & --05:09:02.3 & 21.2 &  n/a  &  n/a  \\
            G16$^b$ & 011111~0201 & 06 Jul 2001 & 02:26:34.2 & --05:09:03.1 &  3.9 &  n/a  &  n/a  \\
            G16$^b$ & 011111~0701 & 14 Aug 2001 & 02:26:35.2 & --05:08:46.6 & 23.7 &  n/a  &  n/a  \\
            G17 & 011111~0301 & 03 Jul 2001 & 02:25:14.3 & --05:09:08.4 & 22.5 &  n/a  &  n/a  \\
            G18 & 011111~0401 & 03 Jul 2001 & 02:23:54.1 & --05:09:09.7 & 28.1 &  n/a  &  n/a  \\
            G19 & 011111~0501 & 04 Jul 2001 & 02:22:34.0 & --05:09:02.1 & 23.7 &  n/a  &  n/a  \\
            \noalign{\smallskip}
            \hline
         \end{tabular}
\begin{list}{}{}
\item[a] pointing G12 has not been used because of the high background level (see text)
\item[b] pointing G16 has been repeated since the first instance was curtailed
\item[c] the reported max exposure is the highest value in the exposure maps of the
         individual cameras, i.e. net of soft proton flares
\item[d] the RA and DEC shifts are the astrometric correction applied (see text) for
         the fields which overlap the VVDS area and for which optical identification
         has been performed
\item[e] more details available through our ancillary web page \url{
http://cosmos.mi.iasf.cnr.it/~lssadmin/Website/LSS/Anc/xmdsepic.html}
\end{list}
\end{center}
   \end{table*}

%---> fig. 1 here
%______________________________________________ figure sample  fig.1
% use begin-end figure for one column, figure* for two columns
%

%__________________________________________________________________

\subsection{Generation of the X-ray source list \label{SecXSL}}

The reduction of the XMDS data has been performed in Milan using 
a streamlined version of the pipeline 
developed by \citet{2002ApJ...564..190B} for the HELLAS2XMM survey.
We refer to the latter paper for the formalism and procedure, and
only mention here the main points
and the few differences.

\begin{figure*}
   \centering
   \includegraphics[bb=37 216 545 615,width=16cm,clip]{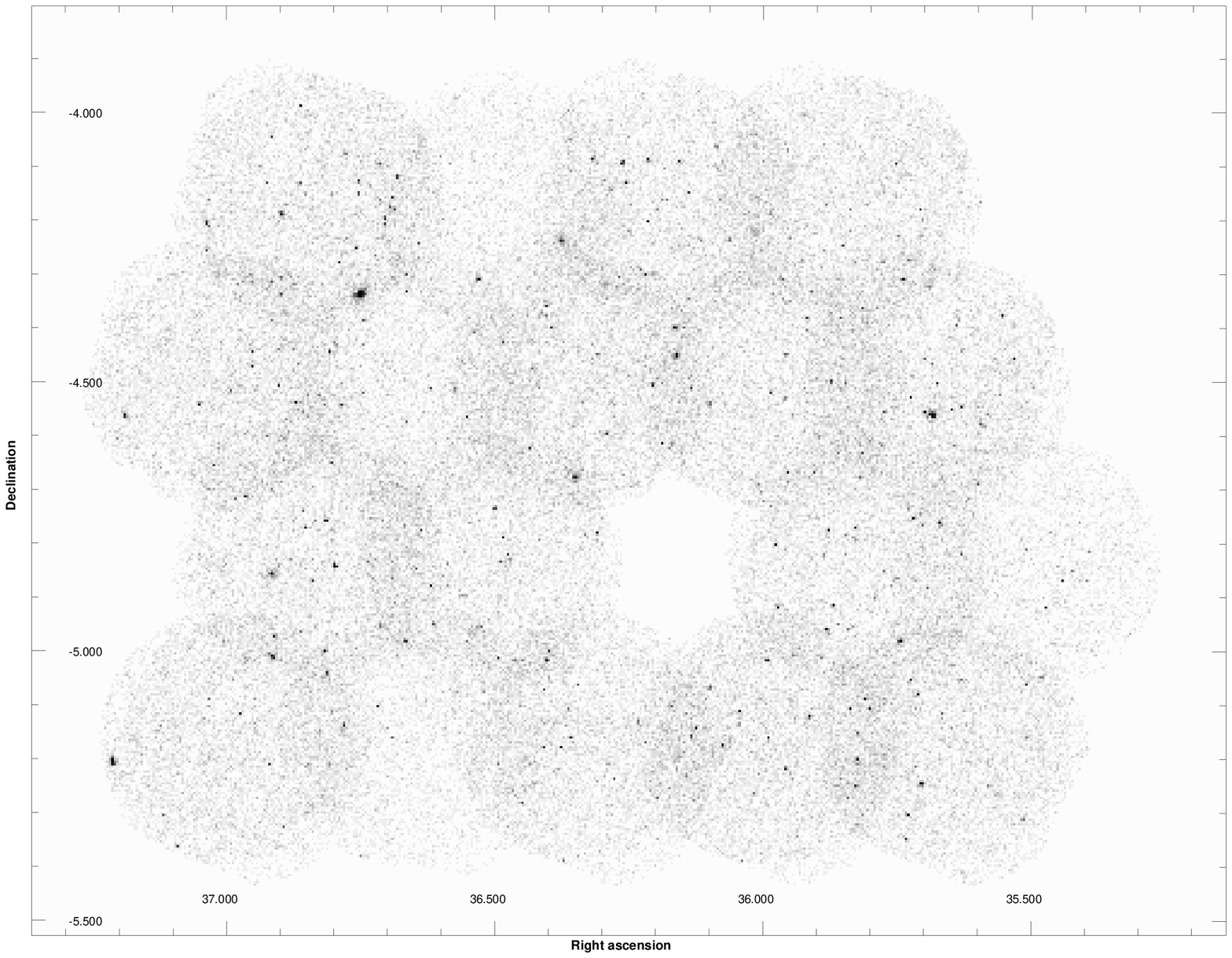}
   \caption{Mosaic of our X-ray images in the total ($0.3-10$ keV) band.
            No exposure map correction has been applied to this figure.
            }
    \label{FigMosaic}
    \end{figure*}

We used the \textit{XMM-Newton} Science Analysis System (XMM-SAS) v5.4.1, whose tasks
{\tt emproc} and {\tt epproc} now provide a reliable identification of
the location of bad pixels. To remove data contaminated by soft proton flares, 
we used {\em fixed} thresholds (on the global background above 10 keV)
of 0.15 counts s$^{-1}$ for MOS
and 0.35 counts s$^{-1}$ for pn : as a result of this  choice, field
G12 could not be analysed because the background was almost constantly above these
thresholds.
A mosaic of our X-ray images is shown in Fig.~\ref{FigMosaic}.

%______________________________________________ figure sample  fig. 2 NEW
% use begin-end figure for one column, figure* for two columns
%

%__________________________________________________________________

As in  \citet{2002ApJ...564..190B}, we ran the source detection (inclusive of
background computation) and
characterization procedure on merged MOS and pn data in {\em five} different
energy bands, 
which we designate as : A ($0.3-0.5$ keV), B ($0.5-2$ keV), C
($2-4.5$ keV), D ($4.5-10$ keV) and CD ($2-10$ keV). 
We note that sources detected in the CD band do not result from
the simple sum of detections in the separate C and D bands,
but from the detection algorithm run on the full $2-10$ keV band,
while quantities referring to the total band
ABCD ($0.3-10$ keV) are obtained by summation of the results in the individual
bands.

We have used the XMM-SAS tasks {\tt eboxdetect} and {\tt emldetect}  to obtain
a list of candidate source positions, but net
counts and associated  parameters are obtained from our own program 
written according to \citet{2002ApJ...564..190B}. In particular
fluxes were computed for a power law spectrum with $\Gamma$=1.7 and
$N_H$=2.61$\times10^{20}$ cm$^{-2}$ (i.e. the average galactic column density
in the XMDS field direction \citep{1990ARA&A..28..215D}), using conversion
factors calculated from response matrices generated consistently with the event
selections used in our pipeline. In particular, as in \citet{2002ApJ...564..190B}, we used a conservative pattern selection of
single and double events for MOS, and only single events for pn.
Since the count rates are derived in regions including a fixed percentage (68\%) of the
expected flux from a point source (whose radius depends on off-axis positions), our fluxes
are correct for point-sources. 
The fact that a source may fall close to (or on) an inter-CCD gap is taken
into account in flux computation by proper usage of the EPIC calibration files,
and by the usage of exposure maps in the measurement of the count rate. This
should be an acceptable approximation \citep[the same used by][]{2002ApJ...564..190B}.
Anyhow, we record in our database a flag indicating whether a source is close to
a FOV edge or to CCD borders or gaps.

For insertion in our database, we retain only sources that have a chance detection
probability less than 2$\times10^{-4}$ {\it in at least one energy band}.
In the 19 useful pointings of the survey, 
we detected 1322 X-ray sources (including
multiple detections in overlapping regions, see Table 2), satisfying the probability threshold $P < 2 \times
10^{-4}$.
In this work we present subsets chosen with more stringent criteria: (1) all sources
detected with chance probability less than
$2\times10^{-5}$ to derive the logN--logS relationship (section \ref{SecLNLS}); (2)
X-ray sources detected with a signal to noise ratio 
larger than 4 in any one of the energy bands defined above
(where the error $\sigma_S$ on the net number of counts is calculated
using the Poissonian approximation from the gross number of counts $cts_{img}$
\begin{displaymath}
\sigma_S=1+\sqrt{cts_{img}+0.75}
\end{displaymath}
according to \citet{1986ApJ...303..336G} as in \citet{2002ApJ...564..190B})
 to study the
mean X-ray properties of the brighter end of the sample (the $4 \sigma$ sample, section
\ref{SecTot}); (3) all sources in the $4 \sigma$ sample covered by VVDS photometry, for which 
we present the catalogue (section \ref{SecCat}) and we 
discuss the X-ray/optical properties (section \ref{Sec4S}).

The numbers of sources in each sample are reported in Table \ref{Tabsamples}.

%__________________________________________________ One column table
\begin{table}
\begin{center}
         \begin{tabular}{lrrr}
         \hline
Sample                   & N$_{tot}$ & N$_B$ & N$_{CD}$ \\
\hline
$P < 2 \times 10^{-4}$   & 1322      & 1166  & 419 \\
$P < 2 \times 10^{-5}$   & 1129      & 1028  & 328 \\
$4 \sigma$ (detections)  & 612       & 591   & 158 \\
%\hline
$4 \sigma$ (independent) & 536       & 518   & 143 \\
$4 \sigma$ VVDS          & 286       & 278   & 73  \\
\hline
\end{tabular}
\caption{Total number of sources and sources detected in the B and CD bands respectively for the
samples presented in the paper. For the $P < 2 \times 10^{-4}$, $P < 2 \times 10^{-5}$ and the
first line of the $4 \sigma$ sample, numbers refer to {\t detections} (i.e. include multiple
detections of the same source in different fields), while for the second line of the $4 \sigma$
sample and for the $4 \sigma$ VVDS sample numbers refer to {\it independent} sources. }
         \label{Tabsamples}
\end{center}
\end{table}

Note that in the present version we have analysed each pointing independently,
in spite of the fact that the FOVs of adjacent pointings overlap. The same
object could therefore be detected in more than one field. This
fact is recognized only {\em a posteriori} (see section \ref{SecTot}). In what follows we will
always specify whether the sample we consider includes multiple detections of the same source.

An independent pipeline is being developed in Saclay for the
analysis of the entire XMM-LSS, using a wavelet technique and therefore best
suited for extended sources (\citet{pacaud05}, see a preliminary account in
\citet{pierre04})
and will be used as a basis for the future complete XMM-LSS catalogue.

\section{The logN--logS relationship \label{SecLNLS}}

We computed the logN--logS distributions
in the $0.5 - 2$ and $2 - 10$ keV band as follows.

We have considered all sources with a detection probability in the
$0.5 - 2$ and $2 -10$ keV bands below $P = 2\times 10^{-5}$. This is well below
our acceptance threshold of $2\times 10^{-4}$, ensuring very few spurious sources in the sample while giving
us a large dataset to compute the logN--logS.
 
\begin{figure}
   \centering
   \includegraphics[width=8.5cm]{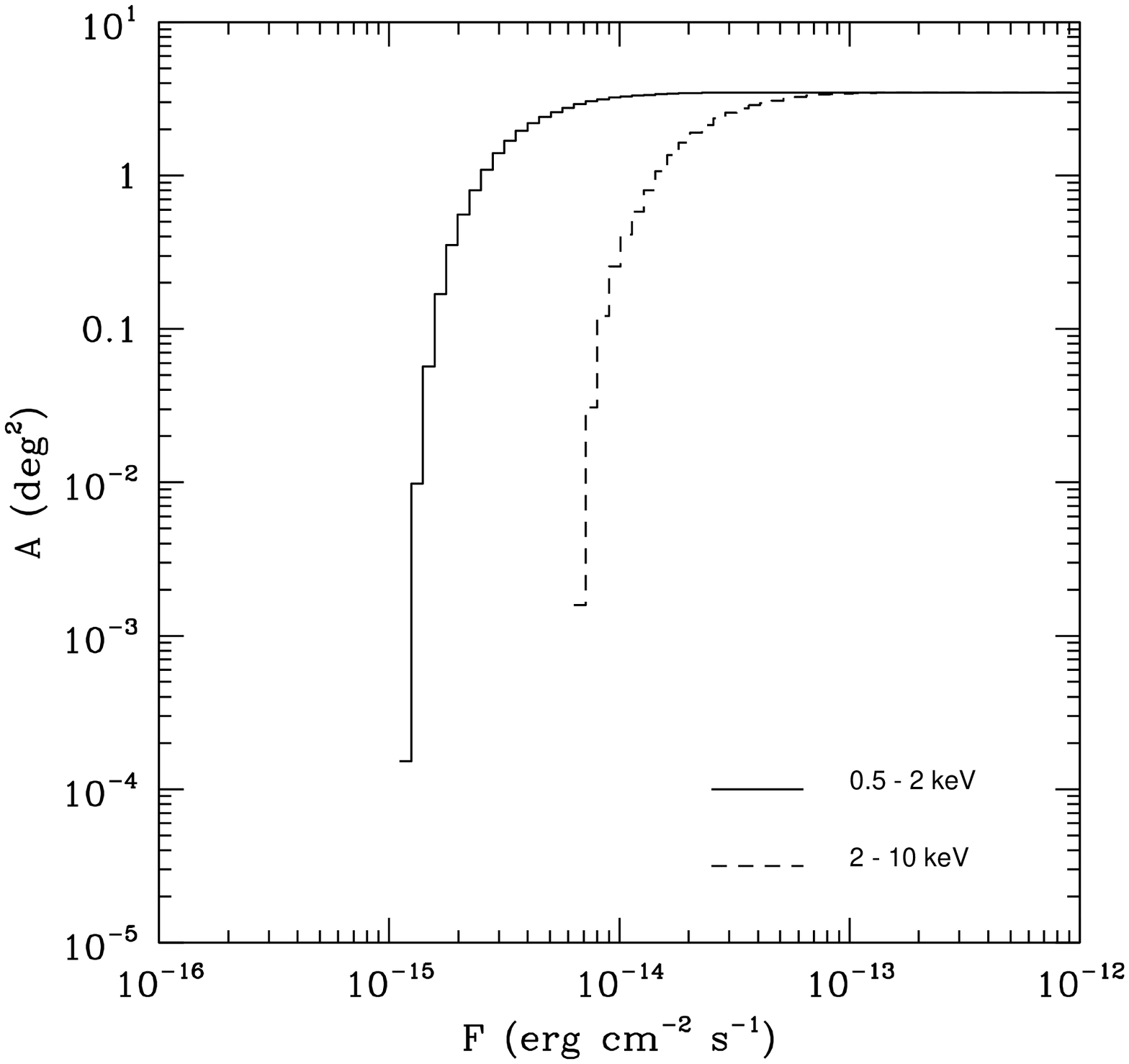}
   \caption{Sky coverage for sources detected at $P < 2 \times 10^{-5}$ in the
   B band (solid line) and in the CD band (dashed line) computed
   considering all fields as independent (see text).}
    \label{FigSkyCov}
    \end{figure}

We have generated a flux limit map for each X-ray field
and energy band, which contains the faintest flux at which a source can be detected
at the assumed level of significance above the local background. From these we have computed the sky coverage
plotted in Fig.~\ref{FigSkyCov}.   This considers all fields as independent areas, 
consistently with our detection procedure (see section \ref{SecXSL}), so the full sky coverage
is the sum of all areas surveyed.

Our faintest flux is $\sim 10^{-15}$ erg cm$^{-2}$
s$^{-1}$ in the $0.5 - 2$ keV band and $\sim 7 \times 10^{-15}$ erg cm$^{-2}$
s$^{-1}$ in the $2 - 10$ keV band.

%--> fig. 3 here
%______________________________________________ figure sample  fig. 3
% use begin-end figure for one column, figure* for two columns
%
%__________________________________________________________________
%--> fig. 4 here
%______________________________________________ figure sample  fig. 4
   \begin{figure*}
   \centering
%  figure with 4 panels  in 2x2 arrangement
   \includegraphics[width=8.5cm]{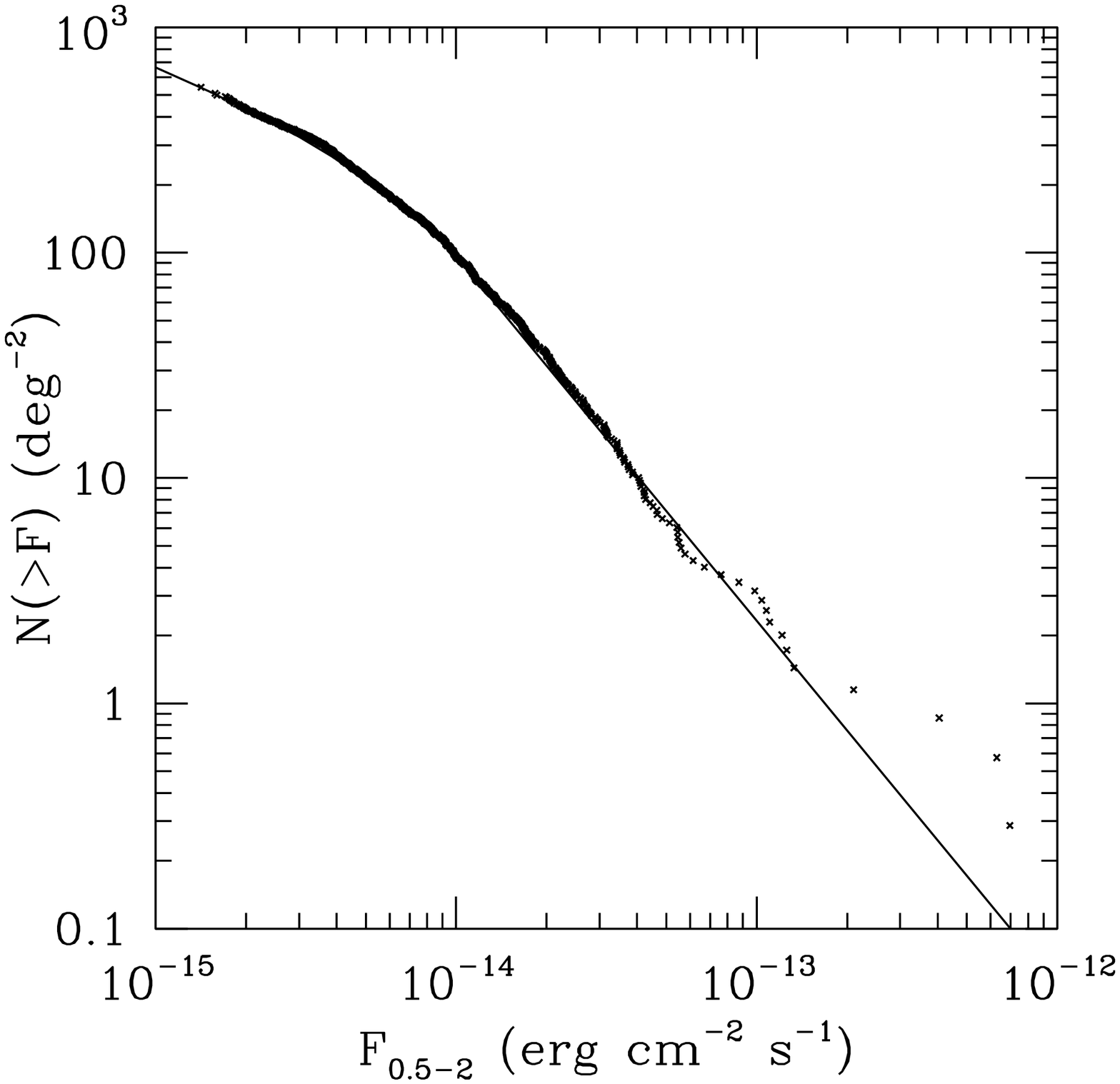}
   \includegraphics[width=8.5cm]{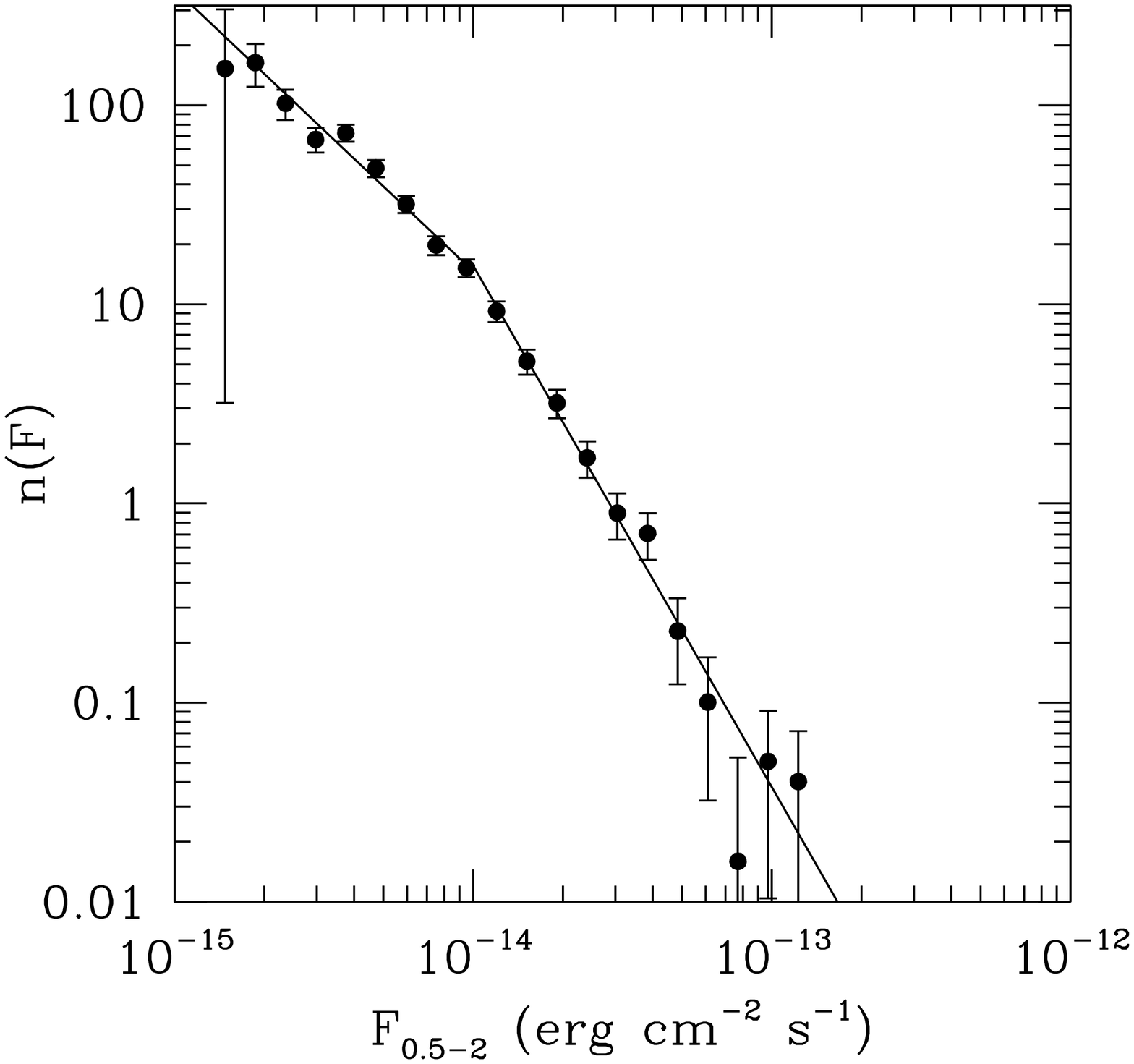}
   \includegraphics[width=8.5cm]{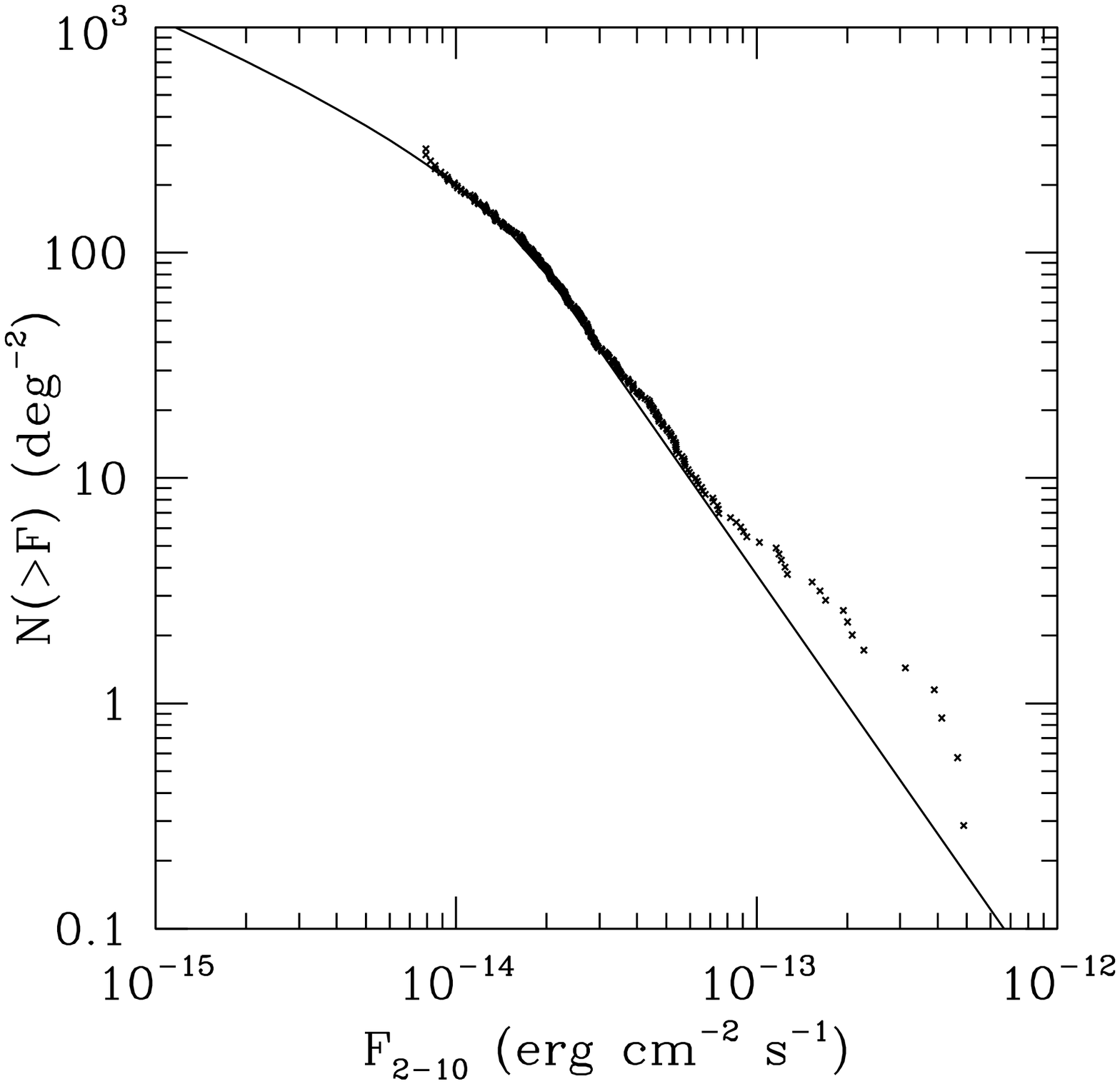}
   \includegraphics[width=8.5cm]{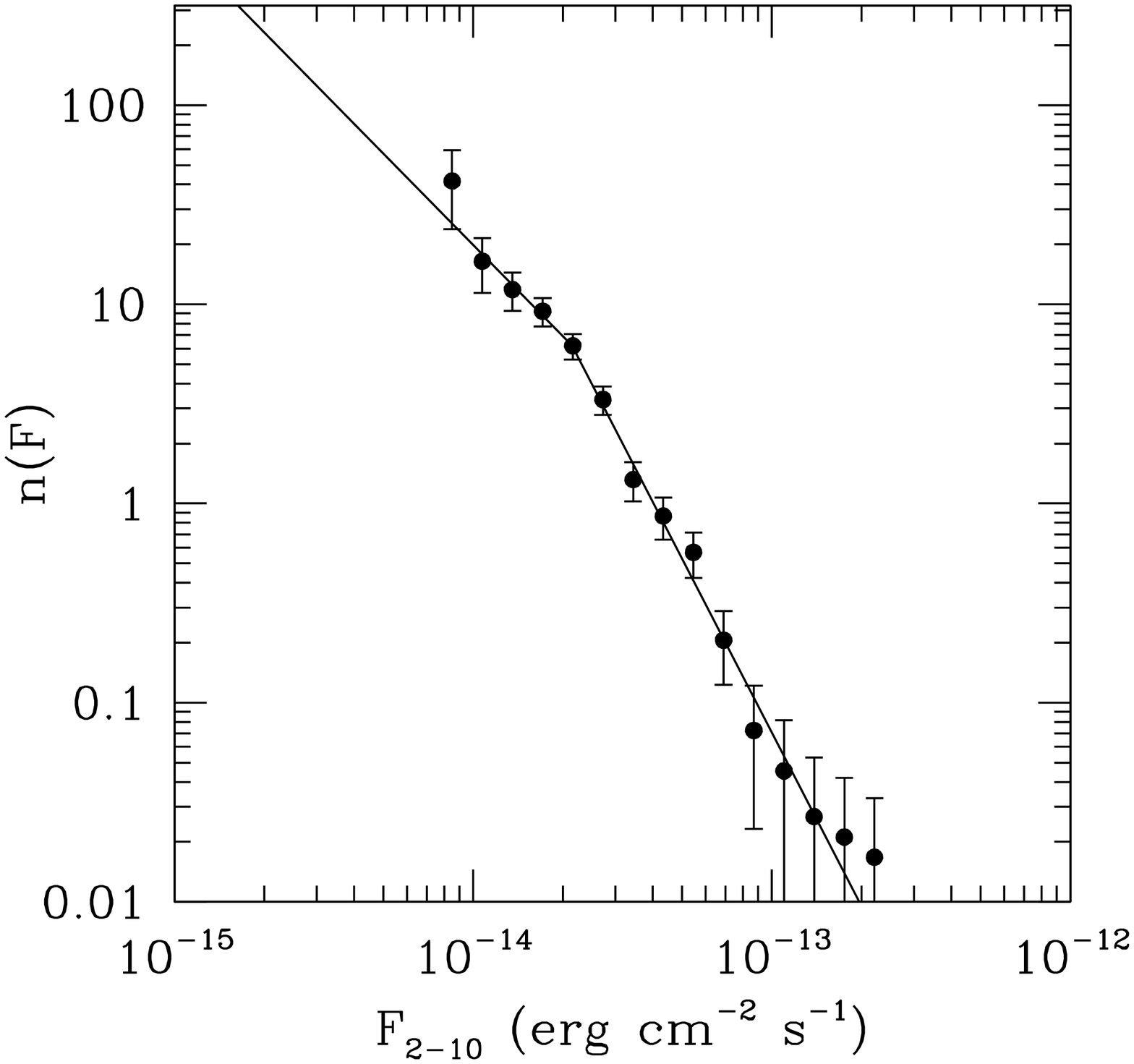}
   \caption{LogN--logS of sources detected at $P < 2 \times 10^{-5}$ in
   band B in the cumulative (upper left panel) and differential (upper right
   panel) form and the same in band CD (lower left and right panels). The units of n(F) are
   number per $10^{-15}$ erg cm$^{-2}$ s$^{-1}$ and deg$^{-2}$ . Our best fits are reported
   as solid lines on the differential plots and have also been converted to the
   cumulative form. 
   }
    \label{FigLNLS}
    \end{figure*}
% use begin-end figure for one column, figure* for two columns
%
%__________________________________________________________________
%
In Fig.~\ref{FigLNLS} we plot the cumulative logN--logS in the $0.5 - 2$ keV
band (upper left panel, 1028 sources) and in the $2 - 10$ keV band (lower left panel, 328
sources). The plots relative to individual fields have also been produced and
inspected and do not show significant deviations from the combined
distribution.

The shapes of the (cumulative) logN--logS are clearly curved and cannot be represented
by a single power law. 
A fit to the cumulative distribution does not allow a meaningful computation of
confidence limits on the parameters, because points and errors are not 
independent \citep{1973ApJ...183....1M,1970ApJ...162..405C}. Therefore
we computed the differential logN--logS 
by binning (in logarithmic space) the
number $N(F)$ of sources with flux $F$ into flux bins of width $\Delta F_i$
(i.e. between fluxes $F_{jmin}$ and $F_{jmax}$), then computing the
average sky coverage $A_i$ in the bin. The $i$-th bin of the differential curve $n(F)$
(shown in the right panels of Fig. \ref{FigLNLS})
is given by
\begin{displaymath}
n(F)_i = \sum_{F_{jmin}}^{F_{jmax}} \frac{\mbox{N}(F_j)}{\Delta F_i A_i}
\end{displaymath}
A statistical error due to the Poissonian error on the number of sources is assigned to each
bin.
Again, we confirm that a single power law does not give a good fit.
Therefore we used a broken power law.

For the $0.5 - 2$ keV band, the best fit is
\begin{displaymath}
 n(F) = \begin{array}{ll}
 6.515 \times 10^3 \times F^{-2.62}_{15} & F_{15} > 10.58 \\
 384.2 \times F^{-1.42}_{15} & F_{15} < 10.58 \\
 \end{array}  
\end{displaymath}
All fluxes $F_{15}$ are normalized to $10^{-15}$ erg cm$^{-2}$ s$^{-1}$. 90\% confidence limits on slopes
are $2.62^{+0.25}_{-0.22}$ and $1.42^{+0.14}_{-0.15}$, while the break position is in the range
$1.06^{+0.30}_{-0.22} \times10^{-14}$ erg cm$^{-2}$ s$^{-1}$.

For the $2 - 10$ keV band we find
\begin{displaymath}
n(F) = 4.483 \times 10^4 \times F^{-2.91}_{15}
\end{displaymath}
for $F_{15} > 21.4$, with 90\% errors on slope
$2.91^{+0.45}_{-0.30}$ and break at
$2.14^{+0.81}_{-0.54} \times10^{-14}$ erg cm$^{-2}$ s$^{-1}$.
At lower fluxes the slope is not well constrained: the
best fit is 1.53, but the confidence interval ranges between 0.37 and 2.04.

In Fig. \ref{FigLNLS} we show the fit results both in the differential and the
cumulative forms for a better comparison with the literature that gives sometimes cumulative 
\citep[e.g.][]{2002ApJ...564..190B,2003ApJ...588..696M}, sometimes differential plots 
\citep[e.g.][]{2004AJ....128.1501Y,2003ApJ...596..944H}. 
Given the uncertainties in the fits, we prefer the comparison with the
actual data where possible.

Our probability threshold $P < 2\times 10^{-5}$ is one of those used in
\citet{2002ApJ...564..190B}. Since our
procedure is very similar to theirs and the flux range is nearly the same, the
logN--logS should in principle be almost identical. We find consistency (within errors) with the
integral HELLAS2XMM distributions for the lower fluxes: $F_{0.5-2} \la 3 \times 10^{-14}$ erg cm$^{-2}$
s$^{-1}$ and $F_{2-10} \la 2 \times 10^{-14}$ erg cm$^{-2}$ s$^{-1}$, while at
higher fluxes we are nominally lower, but, taking into account the purely statistical error on
the number of sources, we are consistent with the lower envelope of the
\citet{2002ApJ...564..190B} relationship.

A good agreement is also found with the \textit{Chandra} Large Synoptic X-ray 
Survey \citep[CLASXS,][]{2004AJ....128.1501Y} and with the Serendipitous Extragalactic X-ray
Source Identification \citep[SEXSI,][]{2003ApJ...596..944H} program, in the common flux range. We
compared the differential plots and found consistency with the CLASXS survey in the soft
band \citep[no plot in that band is given in][]{2003ApJ...596..944H}. In the hard band we are consistent with the SEXSI survey, but we are systematically lower
than the CLASXS points; this is not surprising, given that \citet{2004AJ....128.1501Y} find that their total counts at
$F_{2-8} \sim 10^{-14}$ erg cm$^{-2}$ s$^{-1}$ are $\sim
70\%$ higher than those of \citet{2003ApJ...588..696M}, who use \citet{2002ApJ...564..190B} data for the
flux range we are interested in. 

The sky location of the XMDS survey has
been intentionally chosen in order to avoid known bright X-ray sources and this
could therefore explain, at least partially, the low density observed in our survey for fluxes above
a few $10^{-14}$ erg cm$^{-2}$ s$^{-1}$. On the other hand, \citet{2004AJ....128.1501Y} interpret
their overdensity as an indication of an underlying large scale structure. We note that the
CLASXS, like the XMDS, is constructed on contiguous pointings, while the HELLAS2XMM and the
SEXSI surveys are serendipitous surveys: it seems reasonable that the number count
difference we find between the different surveys is caused by cosmic variance.
   
\section{X-ray properties of the {\em total XMDS} $4 \sigma$ sample \label{SecTot}}

%--> fig 6 here  now 5
%______________________________________________ figure sample  fig. 6  now 5
   \begin{figure*}
   \centering
%  figure with 2 panels side by side
   \includegraphics[width=8.5cm]{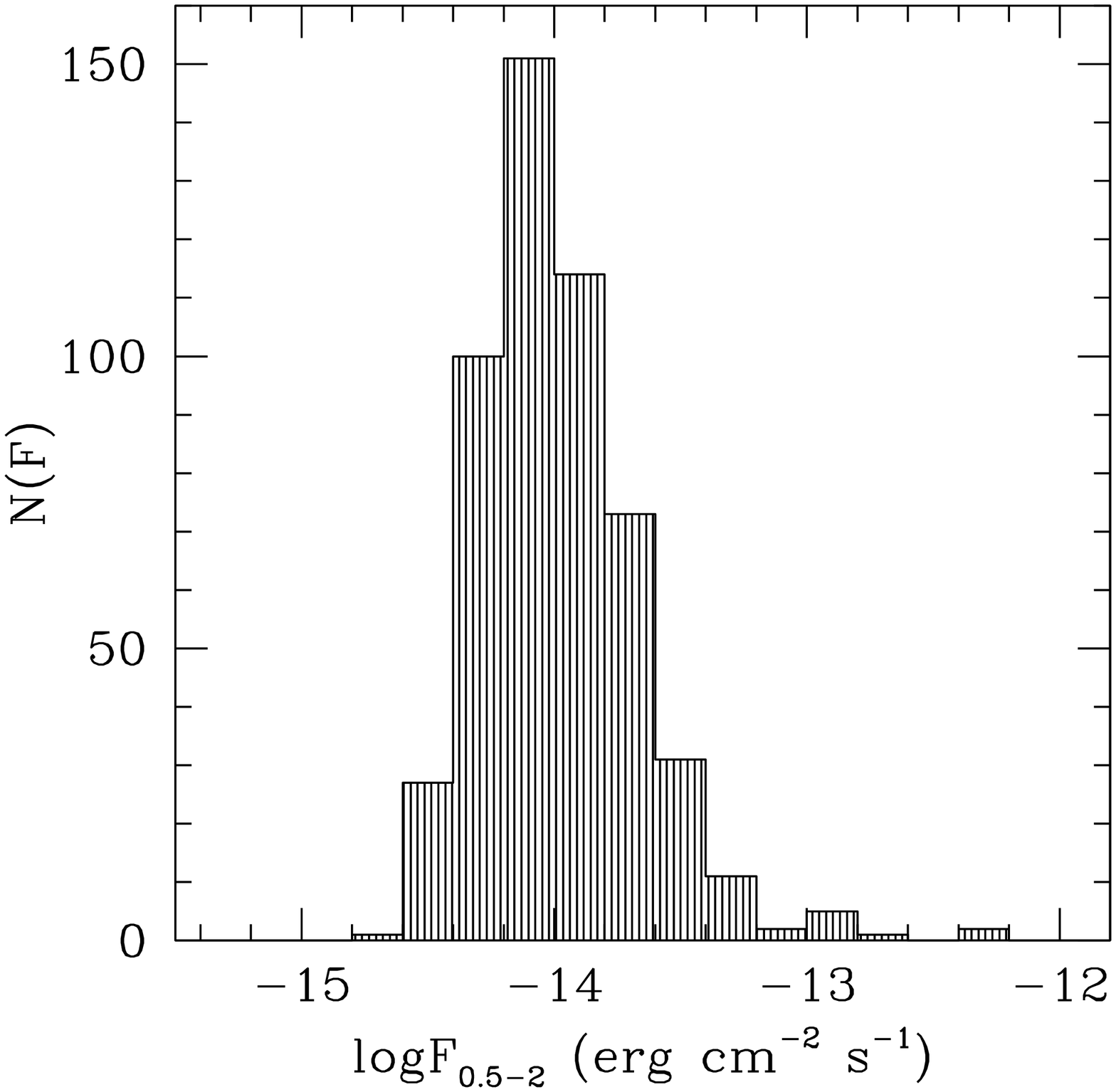}
   \includegraphics[width=8.5cm]{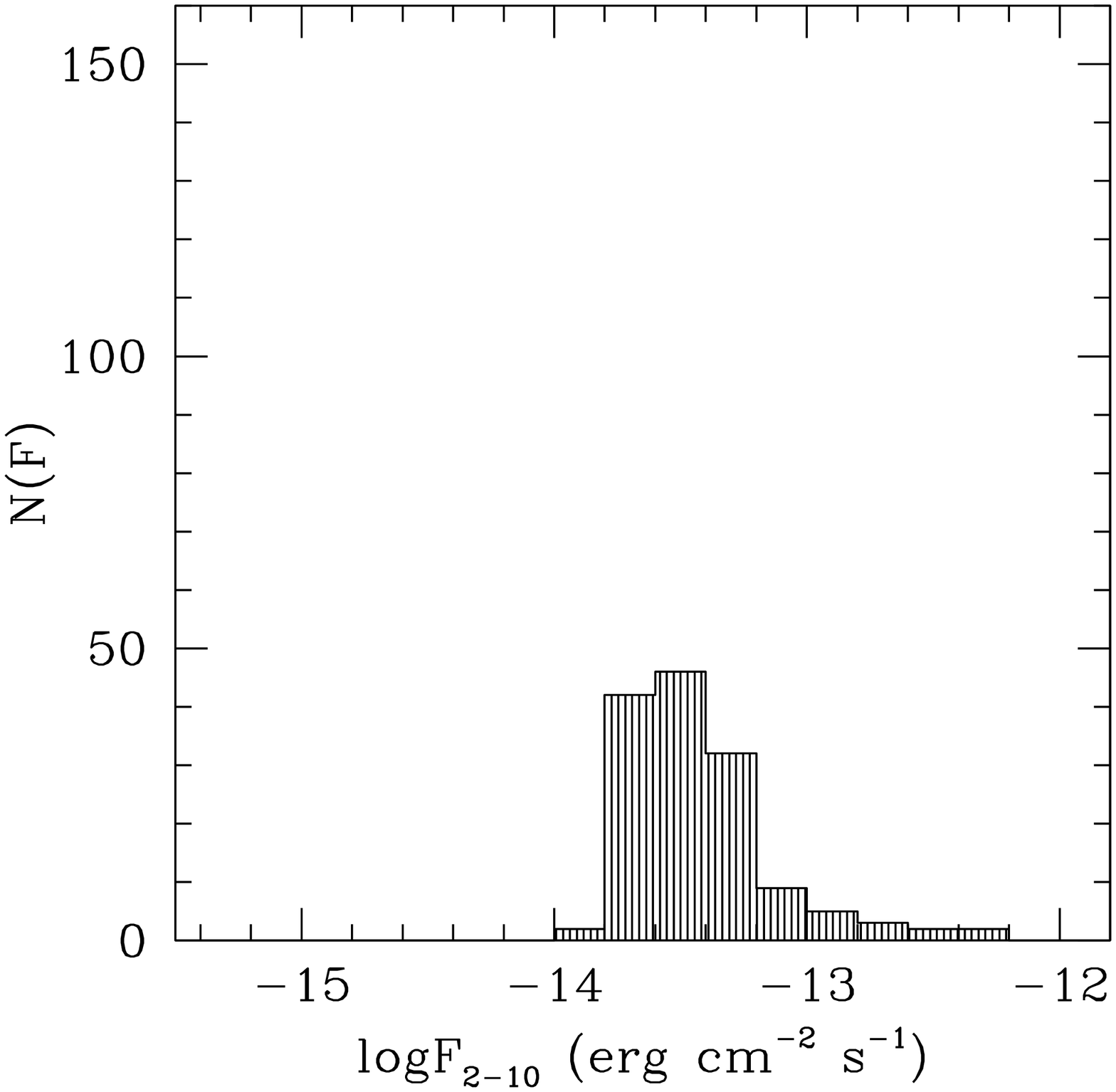}
   \caption{Flux distribution in band B (left panel) and in band CD
            (right panel) for the total $4 \sigma$ sample.
            The hatched area gives the distribution of
            the sources detected at $\geq 4 \sigma$ {\em in the concerned band}.
            }
    \label{FigFlux}
    \end{figure*}
% use begin-end figure for one column, figure* for two columns

%__________________________________________________________________

One of our criteria for the definition
of a sample is the presence of a localized
X-ray count excess  with signal to noise ratio 
(as defined in section \ref{SecXSL})
$\geq 4$ in at least
one of the energy bands (the ``$4 \sigma$ sample''). This 
results in 612 {\em detections}. Taking into account multiple detections
(same source detected in different overlapping fields) the total
number of {\em independent sources} reduces to 536.
All detections in different
pointings with X-ray positions closer than 6\arcsec{} have been automatically considered as multiple
detections of the same source. For
larger separations (up to 18\arcsec{}, which can be considered
as a boundary for suspicious overlaps,
since there are only 2 pairs of sources detected in the {\em same} field which are
closer than such a distance, one at 8.7\arcsec 
~and one at 15.9\arcsec 
), we examined each of the 8 pairs and recovered 2 cases: one is included
in our catalogue and is discussed in Appendix \ref{SecCom}, the other 
has a distance of 6.4\arcsec{}, but one detection is outside the FOV
of one camera.
For all multiple detections
we used the position and derived parameters from the detection that gave
the highest significance, unless the source falls outside the FOV of one
camera, or close to an inter-CCD gap.
We do not detect any significant flux variation in the multiply detected
sources with the possible exception of three, which are however
close to inter-CCD gaps.  The apparent flux decrease in the $0.5-2$ keV band
from the brightest observation in each couple
would be of --35\%, --65\% and --28\%,
with a  significance of $\sim$ 3.5, 3.9 and 4.5 $\sigma$  respectively.

\subsection{Fluxes and hardness ratios \label{SecFlx}}

The flux distribution of the sources in the $4 \sigma$
sample is shown in
Fig.~\ref{FigFlux} for sources detected in the energy 
bands $0.5-2$ (left panel) and $2-10$ keV (right panel) respectively.

We find that 518 sources ($\sim 97\%$)
are included in the sample because of their detection in the
$0.5-2$ keV (B) band, while only 143 ($\sim 27\%$) satisfy
the detection criterion in the $2-10$ keV (CD) band.
Clearly most of the latter are {\it also} detected in the B
band.
We do not have any source included in the sample because of its detection
in the $4.5-10$ keV band \textit{only}, and only 16 hard sources ($\sim 10\%$) are
not detected in the $0.5-2$ keV band.
As indicated by Fig.~\ref{FigFlux}, we find  
\textit{a posteriori} $F_X$ $\sim 2.5 \times 10^{-15}$ erg cm$^{-2}$ s$^{-1}$ as
the flux limit in the B band and $F_X$ $\sim 2 \times 10^{-14}$ erg cm$^{-2}$
s$^{-1}$ in the CD band. 

A spectral analysis of the brighter sources is in progress.  Here we
present the results of the hardness ratio analysis that can be done for
a significantly larger number of sources, and is the
only spectral analysis possible for the fainter sources in the sample.

To this end we note that, even if a source does not satisfy the
detection threshold in a band different from the one that motivated
the inclusion in the sample, we have an estimate of the counts and fluxes in all bands considered,
which we can use to get spectral information. 
Let us define: 

\begin{displaymath}
\mbox{HR} = \frac{CR_{2-10} - CR_{0.5-2}}{CR_{2-10} + CR_{0.5-2}}
\end{displaymath}

\begin{displaymath}
\mbox{HR}_{cb} = \frac{CR_{2-4.5} - CR_{0.5-2}}{CR_{2-4.5} + CR_{0.5-2}}
\end{displaymath}

\begin{displaymath}
\mbox{HR}_{dc} = \frac{CR_{4.5-10} - CR_{2-4.5}}{CR_{4.5-10} + CR_{2-4.5}}
\end{displaymath}
where $CR$ are the net count rates in the given energy band corrected for PSF and
vignetting. We only include sources for which we have a reasonable estimate of the flux
(i.e. a signal-to-noise ratio $\ge 2$) in the $4.5-10$ keV band :
upper limits
in such a band do not carry significant information since they
provide only an upper limit to $HR_{dc}$, which is expected for fainter
sources with the canonical unabsorbed power law spectrum. 
On the contrary upper limits from
the $0.5-2$ keV band are interesting as they provide lower limits to $HR_{cb}$.

We verified that the values obtained from the combined count
rates from different cameras 
are consistent with those computed from single camera data. 

%--> fig. 7 here  now 6
%______________________________________________ figure sample  fig. 7  now 6
   \begin{figure}
   \centering
   \includegraphics[width=8.5cm]{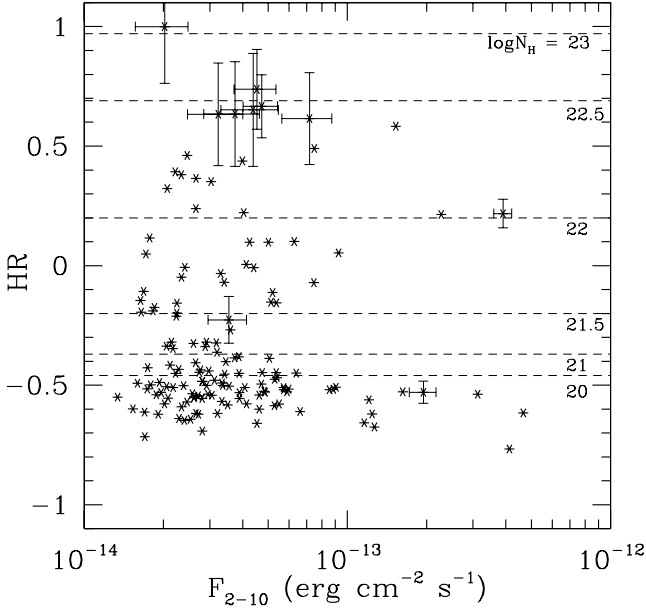}
   \caption{Hardness ratio $HR$ vs $2-10$ keV flux for all sources detected at $\geq 4 \sigma$
   in the CD band (asterisks).
   Only a few typical error bars are shown to avoid figure overcrowding,
   inclusive of all those detected at $\leq 2 \sigma$ in the B band.
   Dashed lines indicate hardness ratios expected for an absorbed power law spectrum
   with energy index $\Gamma = 1.7$ and intrinsic $N_H$ as reported in the
   label. }
    \label{FigHR210}
    \end{figure}
% use begin-end figure for one column, figure* for two columns
%
%__________________________________________________________________

In Fig.~\ref{FigHR210} we plot the main hardness ratio 
$HR$ as a function of the  $2-10$ keV flux. 
Also shown in the figure are dashed lines indicating the hardness
ratios expected for a simple power law model with spectral index 
$\Gamma = 1.7$ and increasing
values of the absorbing column $N_H$. 
This plot shows that $\sim 30\%$ of sources detected
in the CD band have hardness ratios consistent with $N_H > 10^{21.5}$ cm$^{-2}$, for $z=0$.
Note that if sources are at higher redshifts, the horizontal lines for
different values of $N_H$ shift downwards, i.e. the same
hardness ratio corresponds to a higher absorption column.

Within our current sample, we do not find a spectral hardening
 with decreasing  X-ray flux as reported by other authors;
 however our statistics in the hard band is lower than those from other surveys.
On the other hand, it appears from numbers in Table \ref{Tabsamples} that the fraction of
sources in the hard band increases when the sample is larger (from 26\% in the $4 \sigma$ sample
to 32\% in the total $P < 2\times 10^{-4}$ sample). Since a less restrictive threshold implies a
lower flux limit, we can infer that the fraction of hard sources should increase as the flux
decreases. 

%--> fig. 8 here  now 7
%______________________________________________ figure sample  fig. 8  now 7
   \begin{figure*}
   \centering
%  figure with 2 panels side by side
   \includegraphics[width=8.5cm]{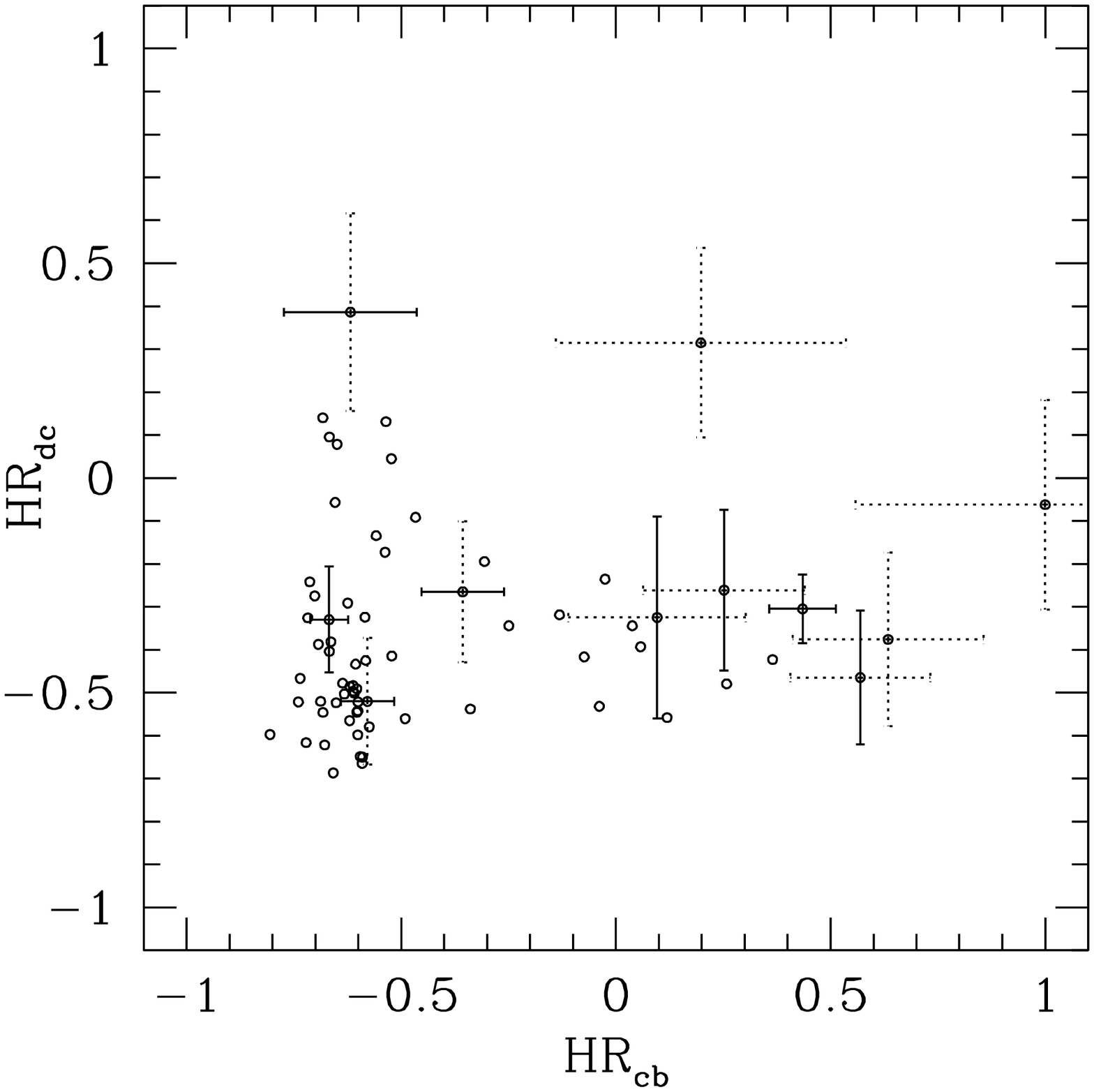}
   \includegraphics[width=8.5cm]{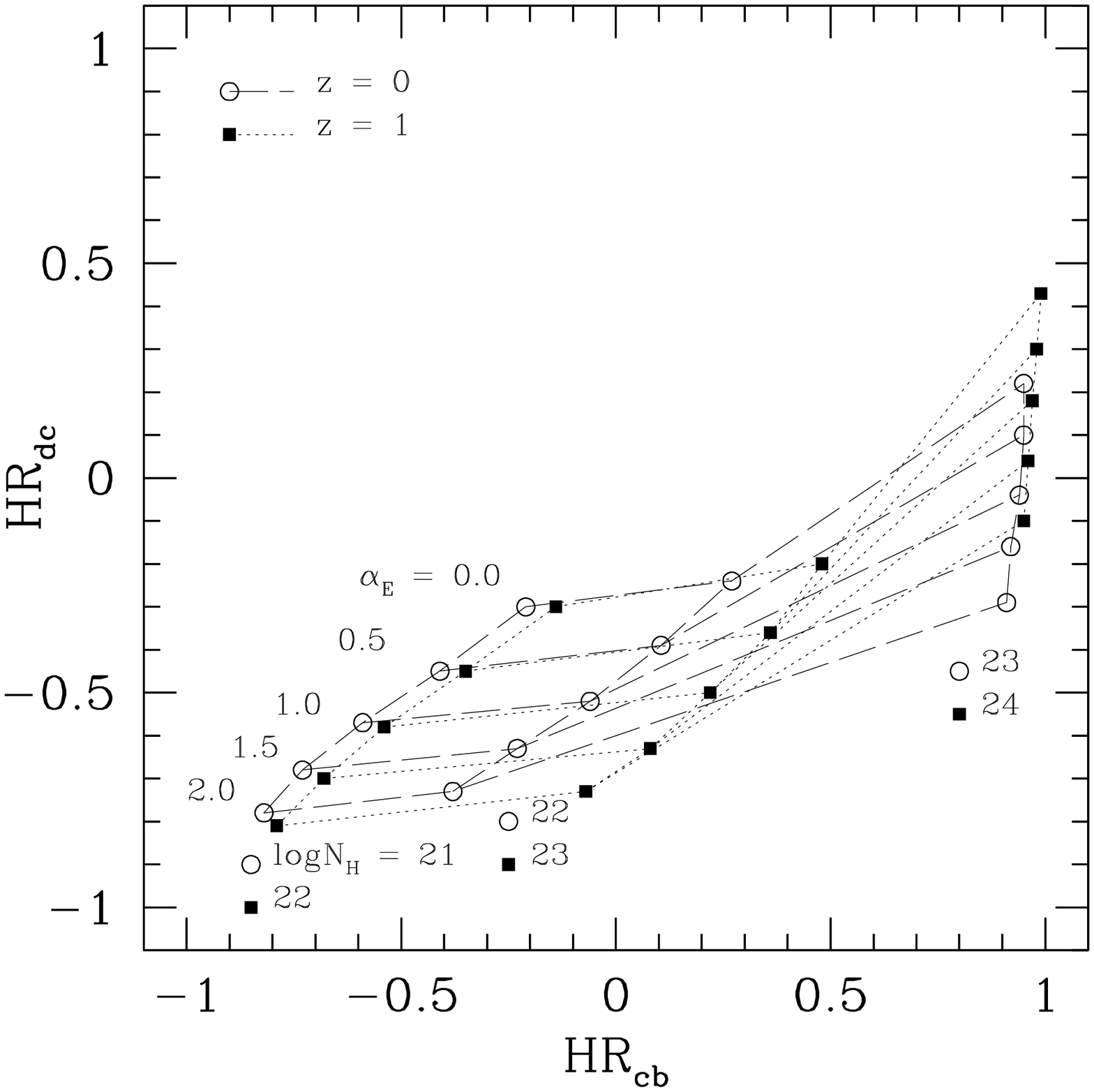}
   \caption{Hardness ratio $HR_{dc}$ vs $HR_{cb}$ for sources in the $\geq 4 \sigma$
      sample (left panel), which are also above $2 \sigma$ in band D.
      Typical error bars 
   are shown (dashed when the significance in the involved bands is below
   $4 \sigma$).
   Grids in right panel represent hardness ratios expected with a simple power law model
   for values of spectral index and $N_H$ reported in the labels at $z = 0$ (empty
   circles and dashed lines) and at $z = 1$ (full squares and dotted lines).}
    \label{FigHRPlot}
    \end{figure*}
% use begin-end figure for one column, figure* for two columns
%
%__________________________________________________________________

Fig.~\ref{FigHRPlot} shows the so-called X-ray colour-colour diagram,
i.e. the plot of the hardness ratio between the harder bands $HR_{dc}$ vs
the hardness ratio between soft and hard bands $HR_{cb}$.
 
Sources with $HR$ consistent with $N_H > 10^{22}$
cm$^{-2}$ in Fig.~\ref{FigHR210} occupy the same locus 
in Fig.~\ref{FigHRPlot} (or $N_H > 10^{23}$ cm$^{-2}$ if we 
consider that sources are at $z \sim 1$). 
This singles them out as promising candidates for more
detailed studies.

In spite of the relatively small number of objects, it is worth noting
that a few sources with low values of $HR_{cb}$ and high
values of $HR_{dc}$ are present: these are not consistent with a
simple power law model.  In particular, their location in 
the diagram at $HR_{dc}> -0.2$ and $HR_{cb}< -0.5$ could suggest a concave
spectral shape, due to the relative lack of signal around 2 keV. 
Such a shape could result 
from the superposition of an unabsorbed component on a flat/absorbed 
power law as often observed in obscured AGNs, where the unabsorbed
component can be due to  scattered or transmitted AGN continuum or
to thermal and non thermal emission associated with starburst
regions.

\section{The XMDS/VVDS $4 \sigma$ sample \label{Sec4S}}

\subsection {The identification procedure \label{SecId}}

For the generation of our sample of candidate identified X-ray sources we
have used the following data sources (database tables and data products).

%--> fig 5 NEW now 8
%______________________________________________ figure sample  fig. 5 NEW  now 8
   \begin{figure*}
   \centering
%  three images in a row
   \includegraphics[bb= 75 201 479 569,width=5.66cm,clip]{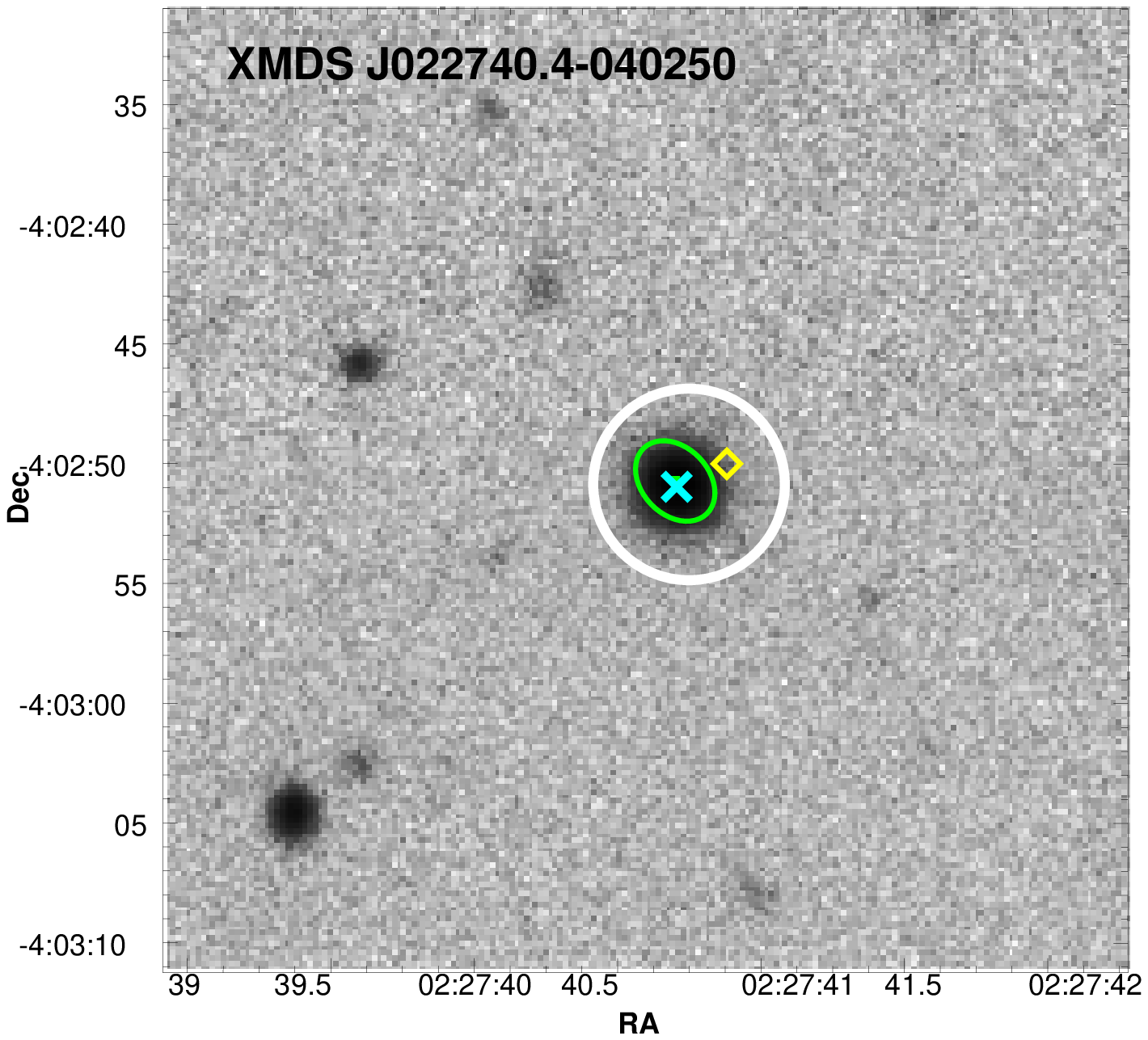}
   \includegraphics[bb= 75 201 479 569,width=5.66cm,clip]{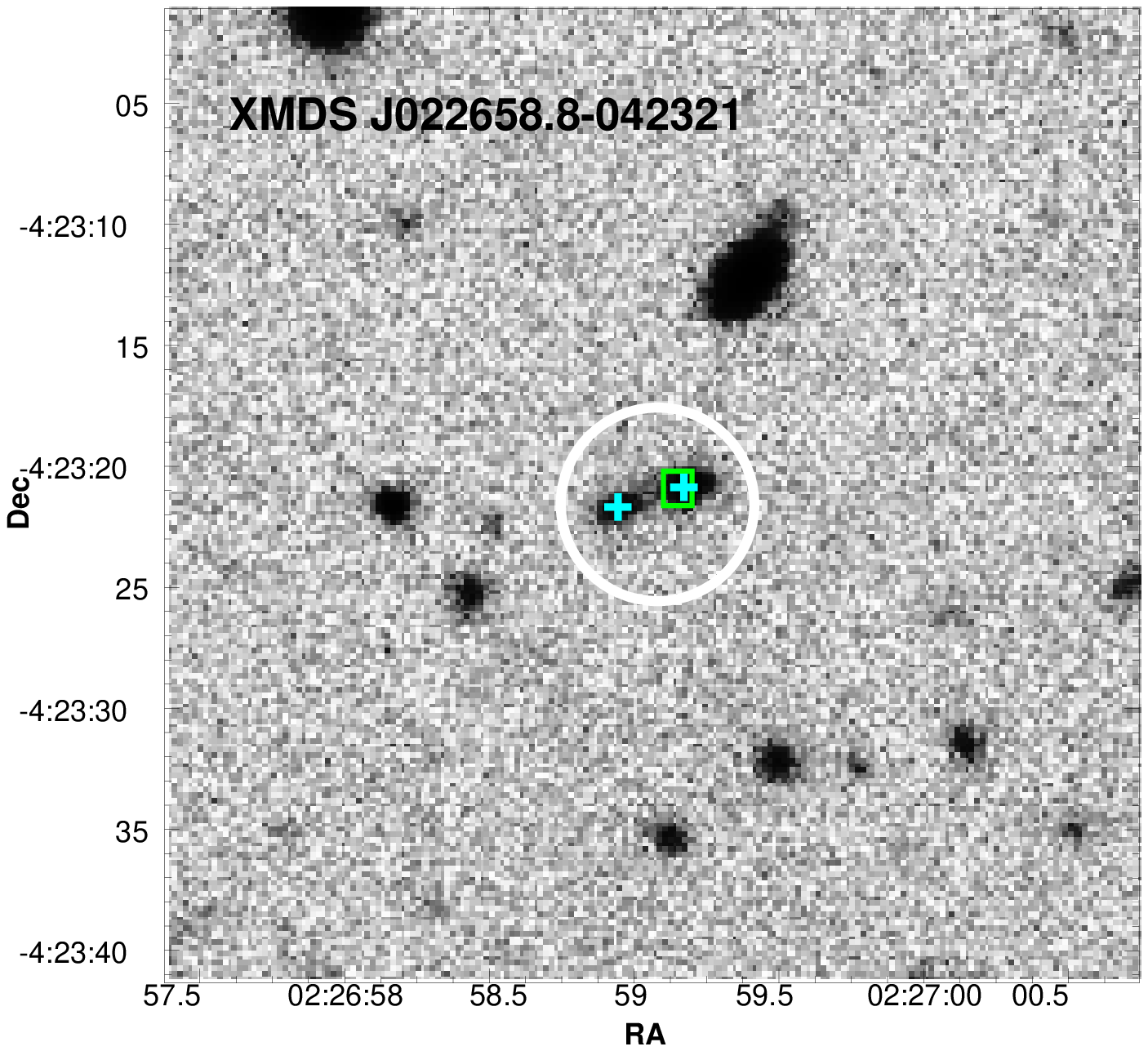}
   \includegraphics[bb= 75 201 479 569,width=5.66cm,clip]{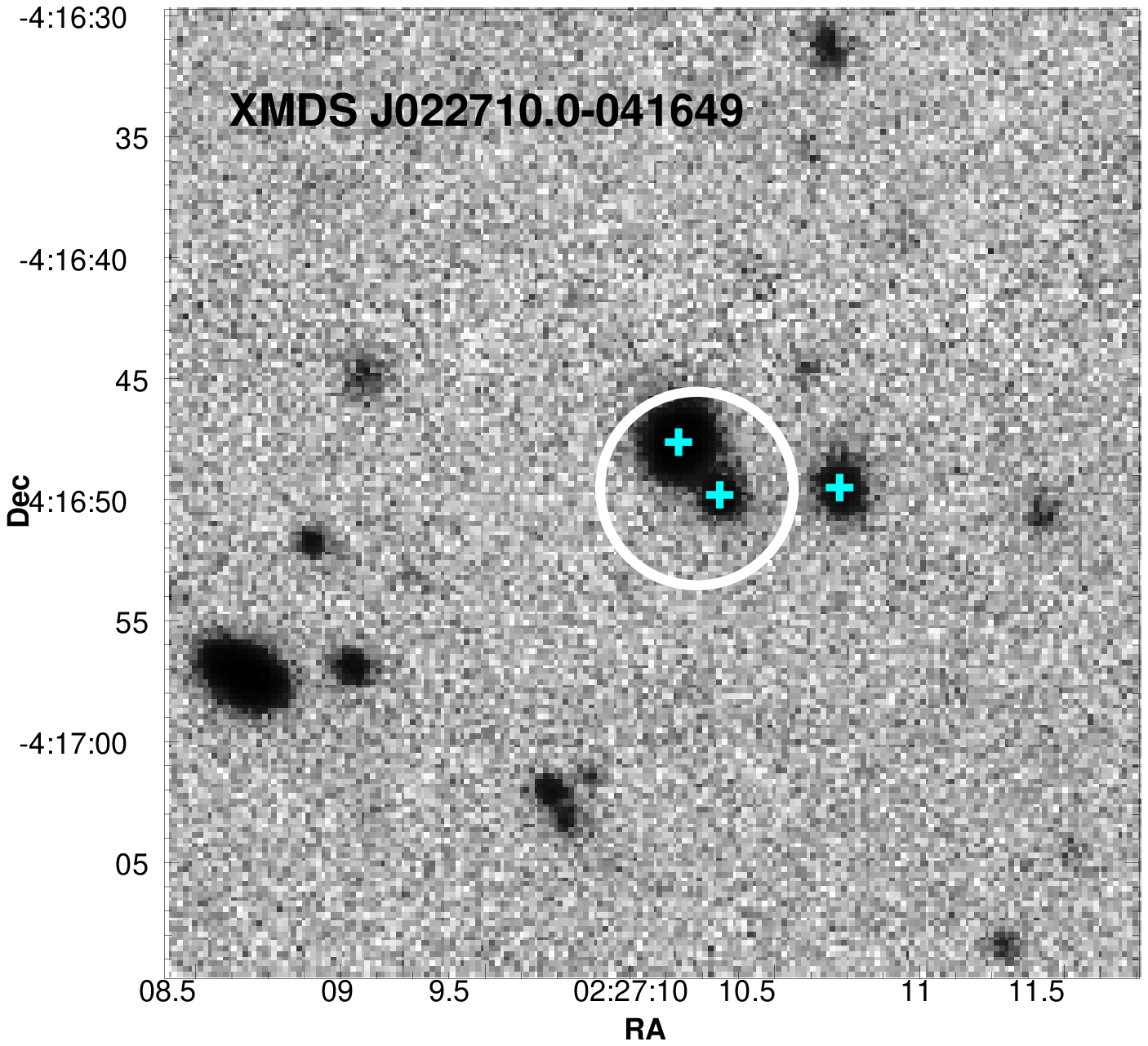}   % 3 in a row
   \caption{Examples from our identification procedure : some representative
   I band finding charts (size 40\arcsec$\times$40\arcsec ) with overlays of
   the X-ray error circles (white thick at 4\arcsec ~radius), the positions of all optical sources within
   6\arcsec ~(cyan X's and crosses respectively for point-like and possibly
   extended objects), the positions of NED catalogue objects (yellow diamonds)
   and the radio source error box (green rectangles) or extent (green ellipses).
   The printed figure is in grayscale : see the online edition for a colour version.
   Three cases are shown : an unique counterpart, an identification preferring a
   radio source, an ambiguous identification.
            }
    \label{FigFC}
    \end{figure*}
% use begin-end figure for one column, figure* for two columns
%

\begin{itemize}

\item
The database table of X-ray sources from
our pipeline.

\item
The optical tables with the VVDS UBVRI(JK) photometry
including both sources with reliable photometric data and
objects whose photometry is flagged as unreliable because the
objects are saturated, masked or close to the borders of the VVDS area.
In addition, we have used the associated finding charts.
Data quality and limiting magnitudes are discussed in
\citet{2003A&A...410...17M} and \citet{2004A&A...417...51R}.

\item
The table with the VIRMOS~1.4GHz radio catalogue
\citep{2003A&A...403..857B} which covers the 1 deg$^2$ VVDS area
at a depth of 80 $\mu$Jy and a resolution of 6\arcsec.

\item
The table with the XMM-LSS radio catalogue 
\citep{2003ApJ...591..640C}, a shallow survey at 74 (4 mJy, 5.6 deg$^2$)
and 325 MHz (275 mJy, 110 deg$^2$) over a larger
area encompassing the XMM-LSS fields.
Occasionally we used the associated radio maps.

\item
The table with the tentative cluster candidates

\item
The NED and SIMBAD database tables, containing pointers to objects
within 20\arcsec ~from an X-ray source.
It should be noted that the
VIRMOS~1.4GHz catalogue has recently been inserted also in SIMBAD, which could
result in a redundant association.

\end{itemize}

The identification process has followed a multi-step procedure, that also involved
the determination of reliable X-ray/optical angular separation criteria for identifications
and the testing of more automated identification procedures for future works. 

First of all, we used our database to create the  sample of X-ray sources
that are within the VVDS area (i.e. for which we can provide a
FITS finding chart, covering a 40\arcsec ~box around the X-ray position, 
that we visually inspected).

For each unique X-ray source we have then conservatively
considered all optical objects within a radius of
6\arcsec ~from the X-ray position \citep[see e.g.][]{2004MNRAS.350..785W}.
This radius is considerably larger
than the nominal SAS ({\tt emldetect}) position error
combined with the 2\arcsec ~\textit{XMM} astrometrical uncertainty.
For such objects we have positions, magnitudes and a crude
optical classification (star-like vs extended) from the VVDS database.

We have identified the optical counterparts 
according 
to the following criteria: a) unique optical counterpart [67 objects]; b) closest and
brightest ($\Delta$mag$>$2) counterpart [166 objects]; c) no optical counterpart down to
I$\sim$25, the magnitude limit of
\citet{2003A&A...410...17M}
[5 objects, of which 3 with no counterpart at all, 2 with
a radio counterpart].
We also used the presence of a radio counterpart as a preference criterion.
The remaining cases [55] do not allow us to assign
an unambiguous optical counterpart until spectroscopic confirmation of the
source is available, so all possible optical counterparts are retained.  
22 cases were flagged as suspect groups, clusters or simply crowded fields,
of which half have however a dominant, unique counterpart.
Examples of finding charts used in our identification procedure are 
shown in Fig.~\ref{FigFC}.

After a first pass with this procedure,
we found evidence in some X-ray fields of a systematic rigid
shift between the position of the X-ray sources and their counterparts
(as high as  3\arcsec ~in field G13). We
profitably used the pairs identified at such stage
to compute (via the SAS task {\tt eposcorr})
a rigid shift correction (see Table~\ref{TabJou}), and inserted in the database two additional columns
with corrected X-ray positions.
No rotational correction was required.
See also section \ref{SecDis}  and Fig.~\ref{FigErrRad}.
After this astrometric correction we repeated our inspection and ranking
procedure.

This time-consuming identification process involved a number of iterations and intermediate steps,
which we have recorded in the database with different
flags and fields (that could identify for example the presence of a visually crowded field, the
cross-correlation with a detection of the same source in a different field, the
source and quality of the optical photometry). These
will remain as part of the record for each source for future automatic database
queries, but are not included in the catalogue attached to this paper.

The final identifications are reported in the catalogue presented here
and have associated numerical codes that describe the quality (rank) of
the identification, as described in point M of Table 3.

As a final step, we have computed the probability of chance coincidence between the
X-ray source and the optical candidates within the given radius according to the
formula by \citet{1986MNRAS.218...31D} used e.g. by \citet{2003A&A...409...65B} or
\citet{2004A&A...418..827M}

\begin{displaymath}
  p = 1 - exp(-\pi~ n(<m)~  r^2 )
\end{displaymath}
where $r$ is the distance between the X-ray source and the concerned optical candidate, and
the density $n(<m)$ of optical objects having I magnitude brighter than the
magnitude $m$ of the candidate counterpart
was computed using the full VVDS catalogue \citep{2003A&A...410...17M}.
A $p<0.01$ criterion (similar to the one used by \citet{2003A&A...409...65B}
and more conservative than the one used by \citet{2004A&A...418..827M},
corresponding in our case e.g. to the probability of finding an I=21 object
at 2\arcsec )
applied to the X-ray/optical pairs produces a list of identifications
very similar to our visually identified sources.
Namely 93\% of our best identifications (rank 0) and 84\% of the rank 1 associations
have $p<0.01$ (and only one of the rank 0 identification has $p\geq0.02$),
while only 21\% of the ambiguous (rank 2) cases have $p<0.01$.

At the present time, spectroscopy is in progress within the framework of different
programs, and only a few of our identifications are spectroscopically confirmed.

\subsection{Optical vs X-ray properties \label{SecDis}}

%--> fig. 9 here
%______________________________________________ figure sample  fig. 9
   \begin{figure}
   \centering
   \includegraphics[width=8.5cm]{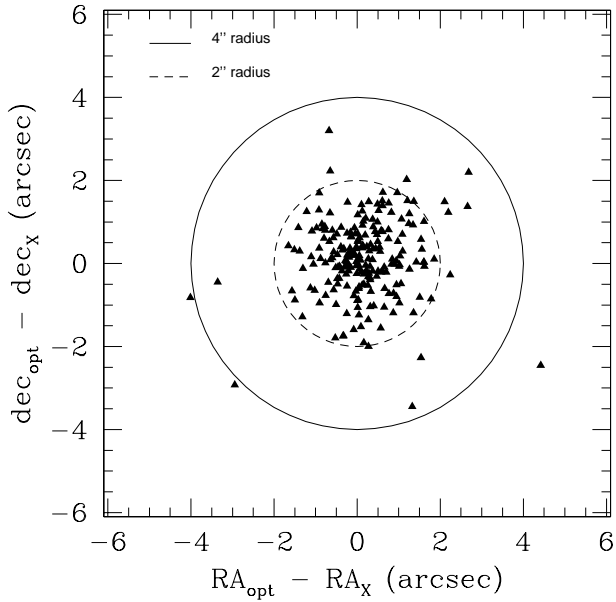}
   \caption{Distance in RA and DEC between the X-ray source position
            (inclusive of astrometric correction) and the
            optical counterpart for well identified sources (ranks 0 and 1
            in the database). Two fiducial radii of 2\arcsec ~and 4\arcsec
            ~are shown.
            }
    \label{FigErrRad}
    \end{figure}
% use begin-end figure for one column, figure* for two columns
%

%__________________________________________________________________

Fig.~\ref{FigErrRad} shows the angular separation between the 
position of the best optical counterpart identified according to the
positional and photometric criteria (as described in section \ref{SecId})
and the position of the X-ray source (after astrometric correction).

One can note the absence of systematic shifts. 
We also verified that there is no correlation between the X-ray to optical
distance and the off-axis position in the field of view, while there is a mild
correlation with count rate, in the sense that sources with a large X-ray to
optical separation are among the weakest.
We note moreover that 93\% of the
optical counterparts are within a 2\arcsec{} radius from the X-ray 
source (99\% within 4\arcsec{}).  We therefore conclude that, after
astrometric corrections have been applied,
the search radius for optical counterparts can be safely reduced to 4\arcsec.

%______________________________________________ figure sample  fig. 10  rev R mag
   \begin{figure}
    \centering
    \includegraphics[width=8.5cm]{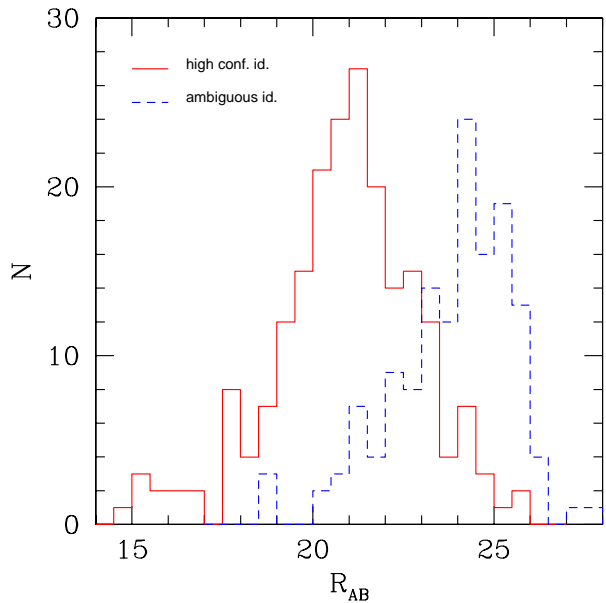}
    \caption{R magnitude distribution for well identified sources (ranks 0 and 1
             in the database, solid histogram) and for optical objects within
             6\arcsec ~from the ambiguously identified sources (rank 2, dashed
             histogram)  }
     \label{FigOpt}
     \end{figure}
% use begin-end figure for one column, figure* for two columns
%

%__________________________________________________________________

%______________________________________________ figure sample  fig. 11 Rev R mag
   \begin{figure}
   \centering
   \includegraphics[width=8.5cm]{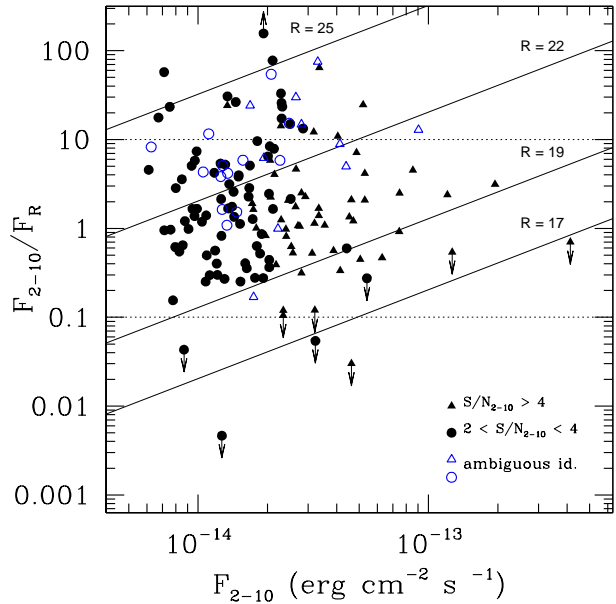}
   \caption{X-ray to optical flux ratio (band CD vs R magnitude) as a function
            of the X-ray flux (in band CD). 
            The sources shown are those in the $4 \sigma$ VVDS sample, detected
            above $2 \sigma$ in band CD.
            Upper limits (on the X-ray
            to optical ratio) correspond to sources photometrically saturated
            (optical magnitude likely underestimated).
            Lower limits corresponds to sources without optical counterpart
            (1 association with a radio source only).
            Open symbols correspond to rank 2 sources
            (ambiguous identification, more than a counterpart, of which the
            brightest is used to derive the optical flux).
            Diagonal lines indicate loci of constant R magnitude
            while horizontal dotted lines mark the region of canonical AGNs 
            ($0.1 <F_X/F_R < 10$)
            }
    \label{FigXOpt}
    \end{figure}
% use begin-end figure for one column, figure* for two columns
%

%__________________________________________________________________
%______________________________________________ figure sample  fig. 12  rev R mag
   \begin{figure}
    \centering
    \includegraphics[width=8.5cm]{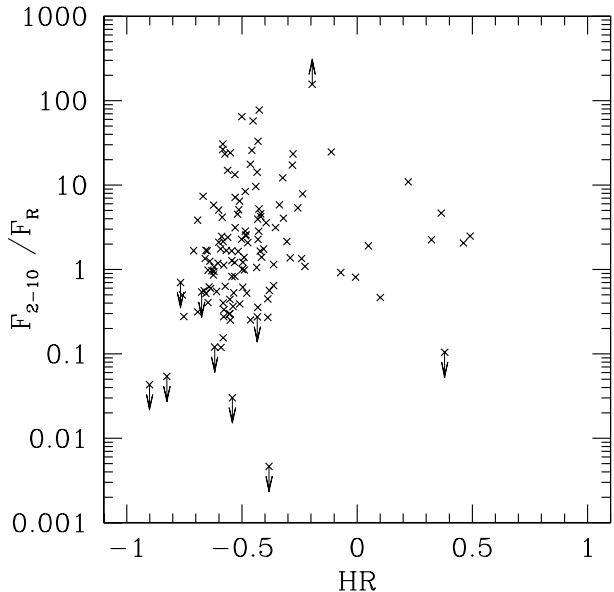}
    \caption{X-ray ($2-10$ keV) to optical flux ratio vs hardness ratio for sources of the
    $4 \sigma$ VVDS sample. Only sources with a significance of at least $2 \sigma$
    in one band (and of course $4 \sigma$ in the other) are shown, provided they have a
    rank 0-1 identification (one case has only a radio counterpart).
    Upper limits are due to "saturated" optical
    counterparts which only yield a lower limit to the optical flux.
   }
     \label{FigXOptHR}
     \end{figure}
% use begin-end figure for one column, figure* for two columns
%

%__________________________________________________________________

Figure \ref{FigOpt} compares the distribution of  optical magnitudes
for the identified sources and for the
ambiguous identifications, i.e. X-ray sources for which there are 2 or more
candidate identifications in the search radius.
As expected, unidentified sources
populate the region of fainter magnitudes where the density of
possible optical counterparts makes chance coincidences more
probable.  Spectroscopy or at least multicolor photometry is
needed to increase the number of identifications and to 
determine the redshifts and types of the candidate counterparts.
Work on the optical and multifrequency information already acquired
within the VIMOS and XMM-LSS consortia is in progress.

We compute the  X-ray to optical flux ratio using our R magnitudes (in
the AB system) according to the formula

\begin{displaymath}
  log(F_X/F_R) = logF_X + 5.51 + R_{AB}/2.5
\end{displaymath}
where the constant derives from the standard conversion of AB magnitudes into
monochromatic flux and from the integration of the monochromatic flux over the
filter bandwidth assuming a flat spectrum.

The X-ray to optical ratio for our identified sources is shown in Fig.~\ref{FigXOpt} as a function of 
the X-ray flux ($2-10$ keV band). We have also included the
ambiguous identifications (marked with different symbols) assuming that
the brighter object is the true counterpart. The latter values should therefore be
considered as lower limits for the X-ray to optical flux ratio:
a few objects that are now in the ``normal AGN"
region could migrate upwards and populate the  obscured AGN region. 

As expected,
most ($\sim 80\%$) of the sources lie in the region of classical AGNs ($0.1 <
F_X/F_R < 10$), $\sim 3\%$ are in the starburst -- normal galaxy
range ($F_X/F_R < 0.1$) and $\sim 17\%$ (including optically blank
fields) in the region expected for optically obscured AGNs
($F_X/F_R > 10$).

The
percentage of high X-ray to optical ratio sources is only slightly lower than in other X-ray surveys of
similar depth \citep[e.g. 20\% in][]{2003A&A...409...79F}.
However, since the $4 \sigma$ VVDS sample is not fully identified and since our identification
procedure favours optically brighter objects, the real fraction of sources with
$F_X/F_R > 10$ could be higher. 

Fig.~\ref{FigXOptHR} shows the X-ray to optical ratio vs
the hardness ratio.
As shown recently by \citet{2004A&A...428..383D}
this diagram is a useful tool
to separate stars from the extragalactic population, and to locate 
different kinds of AGN. Since we have used  the X-ray flux in
the $2-10$ keV band  (for comparison with Fig.~\ref{FigXOpt}),
stars are not included in our diagram as they are not detected in 
the $2-10$ keV band. 
 We verified however that our confirmed stars (via SIMBAD or via preliminary
results of VIMOS spectroscopy)  fall in the expected location in the
$F_{0.5-4.5}/F_R$ vs $HR_{cb}$ plot.  Most of our sources
populate the expected locus of AGNs, with a number of objects with very
high (10-100) X-ray to optical ratios (suggesting optical extinction)
and a few with hard X-ray spectra, pointing to X-ray absorbed nuclei.
In a forthcoming paper we will take advantage of the full multifrequency 
and spectroscopic information to discuss the properties of the populations 
selected by the X-ray survey with respect to those of the optical 
survey. 

\begin{acknowledgements}
This work is based on observations obtained with \textit{XMM-Newton}, an ESA science
mission with instruments and contributions directly funded by ESA Member
States and the USA (NASA).

This work has benefitted from many useful discussions with S. Andreon,
A. Wolter and R. Della Ceca
at INAF-OABrera.

OG, EG and JS would like to acknowledge support by contract Inter-University
Attraction Pole P5/36 (Belgium) and by PRODEX (XMM-LSS)
\end{acknowledgements}

%__________________________________________________ long table landscape
   \begin{table*}
   \caption[]{
            Sample of our X-ray catalogue. We report here only the first few lines of
            the catalogue to show the layout. The full catalogue will be available
            at CDS, as "Online material" or through our database as described in
            Section \ref{SecCat}.
            Comments on peculiar sources are reported in Appendix \ref{SecCom}.
            }
   \label{TabCat}         
  \fontsize{6pt}{8pt}
  \begin{verbatim}
  A |          B          |    C     |     D     |   E  | F  |  G  | H  |  I  |  J  | K L M | N |  O  | P  |  Q  | R  |S T U|              V               |  W
________________________________________________________________________________________________________________________________________________________________
 580 XMDS J022315.4-042558 02 23 15.5 -04 25 58.4   56.7 11.8  22.1 10.5  10.0  20.3  B 1 1  1.4                       1     APMUKS(BJ) B022044.68-043934.9 580
 839 XMDS J022315.3-044320 02 23 15.5 -04 43 18.4   48.2 10.8  22.3  9.8   3.9   9.0  B 0 2  2.1 25.45 0.16 24.24 0.36 1                                    839
 310 XMDS J022316.3-042007 02 23 16.4 -04 20 07.8  121.8 15.7   1.0  7.2  12.2   0.5  B 0 0  1.8                       1                          BD-04 388 310
 846 XMDS J022317.2-044035 02 23 17.4 -04 40 33.5  167.2 18.0  72.4 13.3  17.0  37.3  B 0 1  1.1 18.47 0.00 18.23 0.00 1                                    846
 828 XMDS J022318.8-044616 02 23 19.0 -04 46 13.9  166.7 17.9  53.0 11.5  11.0  17.4  B 0 0  0.2 21.16 0.01 20.77 0.01 0                                    828
 321 XMDS J022319.2-041648 02 23 19.4 -04 16 48.4   55.8 11.4   9.6  8.4   4.1   3.6  B 0 1  1.1 22.62 0.04 21.73 0.03 0                                    321
 820 XMDS J022319.4-044732 02 23 19.6 -04 47 30.4  280.6 22.5 125.6 16.0  17.4  38.7  B 0 0  0.7 19.14 0.01 18.49 0.00 0                                    820
 779 XMDS J022321.8-045740 02 23 22.0 -04 57 38.4   20.1  9.1  98.7 14.7   1.9  47.2 CD 0 0  1.0 20.35 0.01 19.47 0.01 1                                    779
 333 XMDS J022324.9-041452 02 23 25.1 -04 14 52.2   70.5 12.9  28.5  9.8   4.7   9.4  B 0 1  0.6 21.83 0.02 21.31 0.02 0                                    333
 550 XMDS J022326.0-043537 02 23 26.0 -04 35 37.1  115.3 15.3  31.3 10.6  17.7  25.1  B 1 1  1.0 23.16 0.04 22.33 0.04 0                                    550
 315 XMDS J022326.3-041838 02 23 26.5 -04 18 38.5   51.6 11.5   8.9  8.4   4.6   4.0  B 0 0  0.5 24.29 0.08 23.27 0.07 0                                    315
 782 XMDS J022326.3-045708 02 23 26.5 -04 57 05.1  122.8 15.8  62.2 12.3  11.4  29.1  B 0 0  1.5 20.74 0.01 20.49 0.01 1                                    782
 789 XMDS J022329.1-045452 02 23 29.3 -04 54 50.8  245.2 21.2  52.0 11.5  22.0  24.0  B 2 1  1.2 20.25 0.01 19.92 0.01 0                                    789
 789                                                                                      2  3.5 22.13 0.02 21.45 0.02 0                                    789
 571 XMDS J022330.2-043004 02 23 30.3 -04 30 04.2  121.7 15.7  47.0 11.7  14.0  27.6  B 1 0  1.7 20.18 0.01 19.94 0.01 0                                    571
 825 XMDS J022330.6-044633 02 23 30.8 -04 46 31.8  303.2 23.3 100.4 14.8  20.5  33.6  B 0 1  0.8 24.44 0.12 23.37 0.08 0                                    825
 840 XMDS J022330.9-044235 02 23 31.1 -04 42 33.0   59.1 12.4  37.6 10.6   6.3  20.8  B 2 2  5.4 24.77 0.14 24.59 0.24 0                                    840
 780 XMDS J022332.0-045740 02 23 32.2 -04 57 38.7  184.5 18.7  49.5 11.6  18.1  24.6  B 0 0  0.7 19.99 0.01 19.63 0.01 1                                    780
 354 XMDS J022332.2-041022 02 23 32.3 -04 10 22.1  128.4 16.2   9.3  7.6   7.7   2.8  B 0 0  0.3 17.77 0.00 16.93 0.00 2                                    354
 807 XMDS J022333.0-044924 02 23 33.2 -04 49 22.5  143.8 16.7  48.0 11.3  11.2  17.9  B 2 1  0.9 20.60 0.01 20.26 0.01 0                                    807
  \end{verbatim}
  \normalsize
\begin{list}{}{}
\item[A] internal sequence identifier
\item[B] IAU catalogue name of the object (built from uncorrected coordinates as
         provided by the XMM-SAS) ; the X-ray properties of a source, i.e.
         columns from B to L, appear once only on the first catalogue entry for
         such a source. If the X-ray source has further possible counterparts, these columns
         are blank and their value is the same as the first entry with same sequence identifier.
\item[C] astrometrically corrected RA (J2000)
\item[D] astrometrically corrected Declination (J2000)
\item[E] number of net counts in band B
\item[F] error on net counts in band B
\item[G] number of net counts in band CD
\item[H] error on net counts in band CD
\item[I] flux in $10^{-15}$ erg cm$^{-2}$ s$^{-1}$ in band B
\item[J] flux in $10^{-15}$ erg cm$^{-2}$ s$^{-1}$ in band CD
\item[K] band in which S/N ratio is the highest
\item[L] gap flag : can assume a value of 0 (good), 1 (fair), or
          2 (poor) related to the fact that the X-ray
          source falls on inter-CCD gaps in one or more of the EPIC cameras, or close
          to the edge or outside the FOV of one EPIC camera and is based on the
          ratio of the punctual exposures in each camera to the maximum exposure in
          the FOV also for each camera.
          It can also assume the value -1 if the X-ray source is outside the
          FOV of at least one camera (there are no such cases in the $4 \sigma$ sample).
\item[M]  identification rank :  gives the quality of the identification with the counterpart, coded as
          0 optimal e.g. only one counterpart, or brightest and closest;
          1 good; 2 (dubious or ambiguous, e.g. typically more than one possible counterpart listed);
          4 unidentified source
\item[N] distance in arcsec between X-ray source and counterpart
          in the following order of preference :
          the VVDS optical counterpart if any;
          or the radio counterpart in the VIRMOS~1.4GHz catalogue \citep{2003A&A...403..857B};
          or the radio counterpart in the XMM-LSS 325 MHz catalogue \citep{2003ApJ...591..640C};
          or the catalogued NED object; or the catalogued SIMBAD object.
          It is blank if there are no counterparts.
\item[O] R magnitude of counterpart from the VVDS catalogue.
          R and I magnitudes are non-blank when there is a valid measurement in the VVDS
          catalogue. They may be left blank also in case of radio counterparts only, or
          of (usually very bright) objects present {\it only} in external (SIMBAD) catalogues.
\item[P] error on R magnitude
\item[Q] I magnitude of counterpart
\item[R] error on I magnitude
\item[S] photometry reliability flag \citep[derived from][]{2003A&A...410...17M} : 0 good;
         1 flagged as possibly bad, but not saturated;
         2 flagged as bad because saturated 
\item[T] a V indicates presence of a radio counterpart in VIRMOS~1.4GHz catalogue \citep{2003A&A...403..857B}
\item[U] an L indicates presence of a radio counterpart in the XMM-LSS 325 MHz catalogue \citep{2003ApJ...591..640C}
\item[V] name of counterpart, precedence order is NED name, then SIMBAD name including VIRMOS~1.4GHz catalogue
         then name in the radio catalogue by \citet{2003ApJ...591..640C}
\item[W] repetition of identifier in column A
\end{list}
\end{table*}

%__________________________________________________________________

%----------------------------------------------------------------------------
% need to move appendix at the very end to avoid screwup of table numbering 
\twocolumn
\appendix
\section{Comments on individual sources \label{SecCom}}

\begin{itemize}
%Source 580 : 
\item XMDS J022315.4--042558: it is identified with a bright spiral, present also in the NED
database (b\_J=19.75), which however falls in a bad photometric area (close to VVDS border) and
has no valid optical magnitudes.

%Source 310 : 
\item XMDS J022316.3--042007: it has as counterpart the bright (V=11.0) star BD--04 388
which also falls in a bad photometric area close to VVDS border, and has no
valid optical magnitudes ; even if it didn't, it would have been saturated.
The distance reported in the catalogue is overestimated, since the bright
source in the VVDS catalogue is split in two by a saturated spike.

%Source 840 :
\item XMDS J022330.9--044235: it is considered a dubious (rank 2) identification,
since the closest object reported as rank 2 is faint (I=24.5) and at 5.4\arcsec.
It is a borderline case, which could perhaps be a blank field.

%Source 357 : 
\item XMDS J022337.3--040938: it coincides with a bright saturated (I $<$ 14.6) starlike object,
which matches a 7.9 mJy 325 MHz radio source, which is a lobe %348
component of the multiple source CRK2003 J0223.6--0409. %346
Another component %347 
is instead close to the NED object NVSS J022336--040932. %48.
Could it be that the star is a foreground object with respect to the radio source ?
The X-ray optical distance reported in the catalogue is overestimated, since the bright
source in the VVDS catalogue is split in two by a saturated spike.

%Source 560 :
\item XMDS J022403.7--043303: it has as counterpart the 16th magnitude object Markarian 1036, 
which is unfortunately saturated in the VVDS.

%Source 341 :
\item XMDS J022403.9--041326: it falls inside a group of weak uncatalogued objects,
which are affected by the spike of a bright star situated north of them. 
A radio object detected both in the VIRMOS~1.4GHz catalogue and in
our own observations at 325 MHz (CRK2003 J0224.0--0413) %363
is at 13.6\arcsec ~south.
This source has a duplicate detection in a different pointing at 11.7\arcsec{},
and both are inside an unpublished galaxy cluster (Andreon, private communication).

%Source 1067 :
\item XMDS J022413.2--045723: it falls at a blank position in a small group of objects,
unconfirmed yet as a spectroscopic group, of which the closest is at 2.6\arcsec.
Note that this source is in field G18, which is not yet astrometrically corrected.

%Source 542 :
\item XMDS J022445.4--043656: it is considered unidentified (rank 4), falling in a blank field
since the closest object (rejected) is faint (I=25.7) and at 5.0\arcsec.

%Source 45:
\item XMDS J022450.9--042903: its brightest and closest counterpart
is a 21 mag object at
0.5\arcsec{}, however at 2.8\arcsec ~there is a radio source detected in the
VIRMOS~1.4GHz survey as well as in our own 325 MHz data (CRK2003 J0224.8--0429).%401
We kept it as a rank 2 source.

%Source 277 :
\item XMDS J022454.7--040628: its brightest and closest counterpart
is a 21 mag object at 1.1\arcsec,
however there is a fainter object at 2\arcsec ~which has also 1.4 GHz radio emission.

%Source 233:
\item XMDS J022456.0--041725: it appears to match positionally
an uncatalogued enhancement  
in the periphery of the NED
spiral galaxy 2MASX J02245534--0417351 (distance of the nucleus from the X-ray source 14.8\arcsec{}).

%Source 41:
\item XMDS J022524.8--044042: it is associated with a saturated bright elliptical, also
listed in NED (b\_J=20.04) and as a VIRMOS~1.4GHz radio source.

%Source 243 "variable"
\item XMDS J022530.6--041418 is one of the sources mentioned
in section \ref{SecTot} for which the observations in two different pointings give a
marginal flux difference ($3.5 \sigma$). Its counterpart is a saturated object
(nominal VVDS I=16.21), corresponding to the NED elliptical galaxy 2MASX J02253077--0414183.

%Source 178 "variable"
\item XMDS J022544.6--041936 is the other source in the VVDS/$4 \sigma$ catalogue mentioned
in section \ref{SecTot} for which the observations in two different pointings give a
marginal flux difference ($3.9 \sigma$). Its counterpart is also a saturated object
(nominal VVDS I=17.49), corresponding to the NED elliptical galaxy APMUKS(BJ) B022313.84--043304.3.

%Source 211:
\item XMDS J022606.6--040259: it is assigned rank 2 because the likely counterpart is not
too close to the X-ray position and appears to be an unresolved double object.

%Source 488:
\item XMDS J022606.9--043316: its counterpart
should be a background object masked behind
the bright (V=10.37) star BD--05 456 (which is too far to be the counterpart).
For this reason, the line-of-sight association is classified as rank 2.

%Source 1276 :
\item XMDS J022713.6--043910: it has as counterpart the bright (V=10.12) star BD--05 461
not included in the VVDS catalogue; if it were, it would have been saturated.

%Source 444 :
\item XMDS J022714.2--042645: the finding chart 
is saturated by a bright star,
BD--05 480, located southwards with respect to our target, whose position is
partially behind a spike from the star. For this reason there are no
objects listed in the VVDS catalogue, although an object is clearly
visible behind the spike at approximately RA=02:27:14.16 DEC=--4:26:45.0.

%Source 1268:
\item XMDS J022721.7--044153: it has no nearby optical counterparts, but a very good
positional coincidence (0.8\arcsec{}) with a VIRMOS~1.4GHz radio source.
Note that a 17 mag saturated object is however present 6\arcsec ~north.

%Source 476:
\item XMDS J022725.1--043656: it is considered unidentified (rank 4), falling in a blank field
since the closest object (rejected) is faint (I=25.8) and at 4.2\arcsec.
                       
%Source 111:
\item XMDS J022735.6--041317: it corresponds to a plethora of very faint objects all located
inside an extended radio halo, whose central position matches very well the
X-ray position.

%Source 120:
\item XMDS J022735.7--041122: there are two possible optical counterparts 
which
are both located  at the periphery, but inside the nominal extent, of the extended 325 MHz radio source CRK2003 J0227.6--0411 %512
centered 19\arcsec ~to the north-east. A 1.4 GHz point-like source,%992
which is a component of a terribly elongated multiple object,
is instead located in the X-ray error radius between the two optical objects, 
while the NED object NVSS J022736--041117 is located at 9\arcsec ~midway between the
X-ray source and the 325 MHz radio source.

%Source 403:
\item XMDS J022742.1--043607: there are no optical catalogued sources;
an unresolved double source is present at 4\arcsec ~just
outside of the nominal error radius of 
the X-ray source and 
within 1.1\arcsec{} 
there is a 1.4 GHz radio point source.
The finding chart is affected by the spikes of a nearby bright star.

%Source 129:
\item XMDS J022748.2--041013: it has no optical counterpart at all, hence is flagged as rank 4, blank field.

%Source 914:
\item XMDS J022812.5--045717: it is in field G15 which has not yet been astrometrically corrected.

\end{itemize}
Concerning spectral peculiarities, we can use as an indicator the fact that
the energy band with the best signal to noise (column K of the catalogue) is
not band B.
We have only one source ( %575
XMDS J022410.7--042759) which is very soft (detected only in band A), and 10
cases with the best signal to noise in band CD (of these 7 are among the 16
not detected in the softer bands mentioned in Section \ref{SecFlx}).

  \Online
% use xmds1.dat in verbatim mode
% paginate forcing a page break inserting a blank line after sources
% 554, 134 and 419
  \onecolumn
  We report here the complete file with our catalogue. For an explanation of the columns
  refer to the sample table \ref{TabCat} in the main text.
% \fontsize{8pt}{10pt}
  \fontsize{6pt}{8pt}

  \end{document}